\documentclass[%
reprint,
 amsmath,amssymb,
 aps,
 prx,
floatfix,
]{revtex4-2}

\usepackage[utf8]{inputenc}
\usepackage{xcolor}
\usepackage[margin=0.75in]{geometry}
\usepackage{graphicx} 
\usepackage{physics}
\usepackage{amsmath}

\usepackage{siunitx}

\begin{document}

\title{Quantum Computing with Circular Rydberg Atoms}
\author{Sam R. Cohen}
\affiliation{Department of Physics, Princeton University, Princeton, NJ 08544}
\author{Jeff D. Thompson}
\affiliation{Department of Electrical Engineering, Princeton University, Princeton, NJ 08544}
\email{jdthompson@princeton.edu}
\date{\today}

\begin{abstract}
    Rydberg atom arrays are a leading platform for quantum computing and simulation, combining strong interactions with highly coherent operations and flexible geometries. However, the achievable fidelities are limited by the finite lifetime of the Rydberg states, as well as technical imperfections such as atomic motion. In this work, we propose a novel approach to Rydberg atom arrays using long-lived circular Rydberg states in optical traps. Based on the extremely long lifetime of these states, exceeding seconds in cryogenic microwave cavities that suppress radiative transitions, and gate protocols that are robust to finite atomic temperature, we project that arrays of hundreds of circular Rydberg atoms with two-qubit gate errors around $10^{-5}$ can be realized using current technology. This approach combines several key elements, including a quantum nondemolition detection technique for circular Rydberg states, local manipulation using the ponderomotive potential of focused optical beams, a gate protocol using multiple circular levels to encode qubits, and robust dynamical decoupling sequences to suppress unwanted interactions and errors from atomic motion. This represents a significant improvement on the current state-of-the-art in quantum computing and simulation with neutral atoms.
\end{abstract}

\maketitle

In recent years, neutral atom arrays in optical tweezers interacting via Rydberg states have emerged as a leading platform for quantum simulation and quantum computing \cite{Saffman2010,browaeys2020}. They combine several attractive features: flexible experimental geometries, large system sizes, excellent coherence and strong interactions. This has enabled explorations of many-body quantum dynamics \cite{zeiher2017,Bernien2017,Lienhard2018,guardado-sanchez2018,Omran2019} and high-fidelity gates \cite{Isenhower2010,Wilk2010,Jau2015,Levine2019,Graham2019,madjarov2020}.

The fidelity of gates based on the interaction between Rydberg states is fundamentally limited by the finite lifetime of the Rydberg states relative to the achievable operation speed. The lifetime of laser-accessible states with orbital angular momentum $\ell \leq 2$ is 100-200 $\mu$s at room temperature, limited by blackbody radiation, and can be improved to 1 ms in a cryogenic environment \cite{Saffman2010}. Circular Rydberg states with maximal angular momentum $|m| = \ell = n-1$ have longer lifetimes, reaching $\approx 10$ ms at cryogenic temperatures, because they have only a single (microwave-frequency) radiative decay pathway, to the next-highest circular state \cite{Hulet1983}. Furthermore, their lifetime may be significantly extended inside a microwave structure that suppresses the local density of states (LDOS) at this single transition frequency \cite{kleppner1981,Hulet1985}. In principle, radiative lifetimes exceeding 100 seconds can be realized, six orders of magnitude longer than for laser-accessible low-$\ell$ states.

Unfortunately, this increased lifetime does not directly translate to improved gate fidelity within conventional Rydberg gate approaches based on the Rydberg blockade, because of the difficulty of exciting circular Rydberg states with high fidelity \cite{Xia2013}. In a standard blockade gate \cite{Lukin2001}, a control atom in a superposition of two ground states $\alpha \ket{g} + \beta \ket{g'}$ is excited to a Rydberg state $\ket{r}$ conditioned on starting in the state $\ket{g}$, resulting in the state $\alpha \ket{r} + \beta \ket{g'}$. Then, a nearby target atom is driven on the same transition from $\ket{g}$ to $\ket{r}$, but this excitation is blocked if the control atom is already in a Rydberg state, giving rise to an entangling gate. While it is straightforward to drive the transition from $\ket{g}$ to $\ket{r}$ with a laser if $\ket{r}$ is a low-$\ell$ Rydberg state, it is quite difficult if $\ket{r}$ is a circular Rydberg state, because of large angular momentum difference ($\Delta \ell \approx 50$) requires a many-photon process. To date, the highest reported fidelity for exciting a circular Rydberg state from a low-$\ell$ Rydberg state is only 90-95\% \cite{Signoles2017,Teixeira2020}.

\begin{figure}[t]
    \centering
    \includegraphics[width=3.3 in]{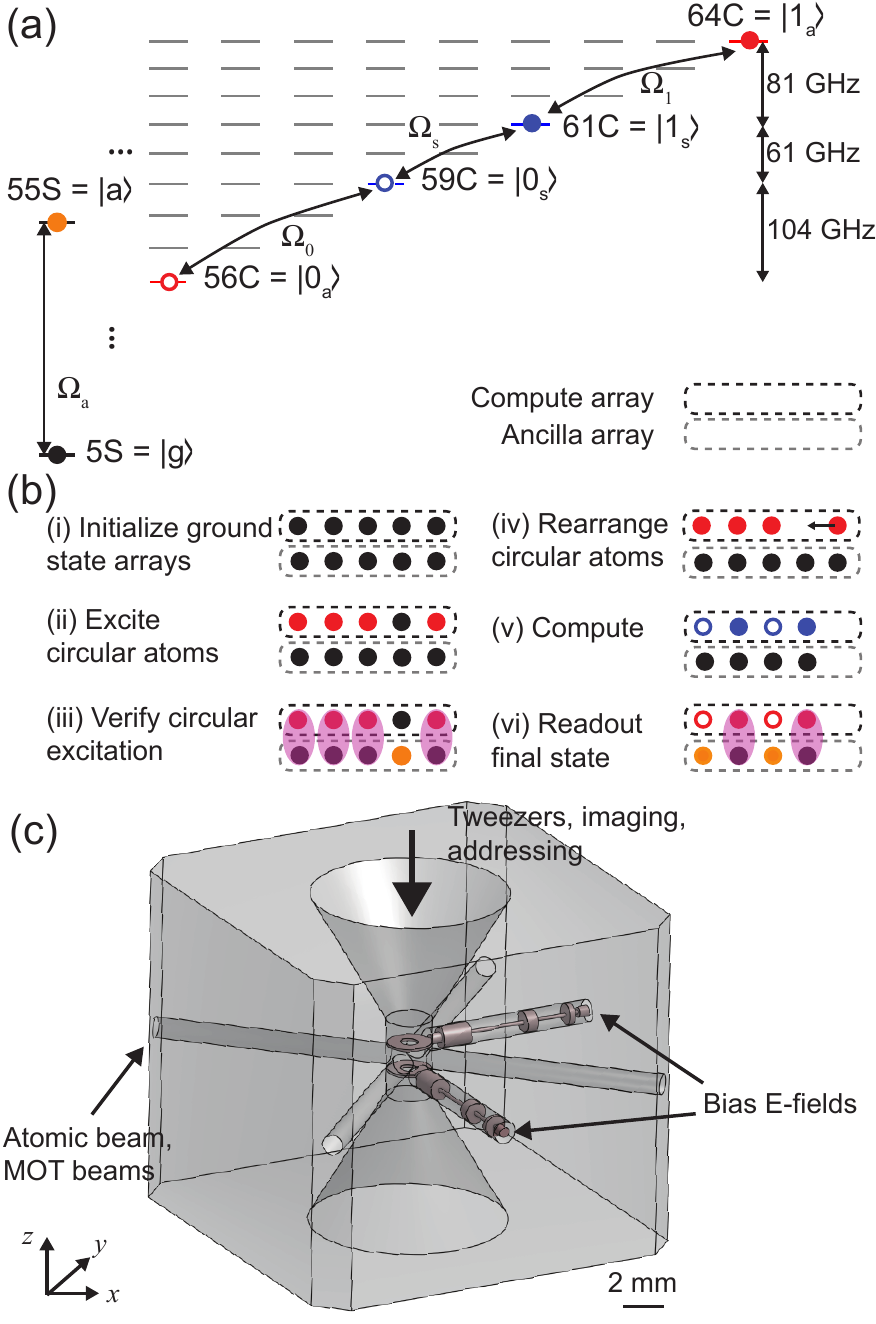}
    \caption{(a) Relevant energy levels and transitions for implementation with Rb. $nC$ denotes the circular state with principal quantum number $n$, while $nS$ is the corresponding $S_{1/2}$ state ($5S$ is the ground state). (b) Schematic diagram of the proposed experimental sequence (Section \ref{sec:overview}). Each circle represents a single atom, with a state color-coded following panel (a).
    (c) Schematic of microwave waveguide used to enhance the circular atom lifetime. The vertical bore ($D=2.2$ mm in the center) provides NA=0.5 optical access for tweezers, imaging and addressing beams, while the side bores allow atoms, cooling light and lattice beams into the structure. Electrodes apply a uniform electric field along $\hat{z}$, and are biased via a reflective stepped-impedance low pass filter. \label{fig:fig1} }
\end{figure}

In this work, we propose an approach that sidesteps this challenge by instead using multiple circular Rydberg states to encode qubits, avoiding the need to repeatedly drive back and forth to ground states. This idea was recently explored in Ref. \cite{Nguyen2018} in the context of quantum simulation. There are several significant novel aspects to our approach that enable the generalization of this idea to programmable quantum computing with individually addressed atoms. First, we propose a waveguide-based microwave structure that enables LDOS suppression while maintaining high-NA optical access for trapping and manipulation. Second, we discuss a rapid, site- and state-resolved nondestructive measurement technique for single circular Rydberg atoms using an ancilla atom array. This enables tweezer-based rearrangement \cite{Kim2016,Endres2016,Barredo2016} to replace defects from imperfect excitation of circular atoms, as well as nondestructive measurements of the circular qubit states. Third, we outline an approach to state-insensitive optical trapping of individual circular Rydberg atoms to reduce motional decoherence of the qubit states. Fourth, we describe a technique for local manipulation using the ponderomotive potential of focused Laguerre-Gauss (LG) beams, enabling site-addressed manipulation of the atoms between the circular Rydberg states used to encode the qubit. Finally, we propose a specific gate protocol using four circular levels (Fig. \ref{fig:fig1}a): a pair of ``storage" levels with weak interactions that can be cancelled by global dynamical decoupling, and a pair of ``active" levels to implement two-qubit gates.

With this approach, we estimate that arrays of more than 200 trapped circular atoms with lifetimes exceeding 3 seconds can be realized. In combination with a projected two-qubit gate duration of $t_\pi \approx 4\,\mu$s, this sets a lifetime limit on the two-qubit gate fidelity of approximately $\mathcal{F} = 1-10^{-6}$. Including realistic experimental parameters and leading sources of error, we estimate that two-qubit gate fidelities $\mathcal{F} > 1 - 10^{-5}$ are achievable, which compares very favorably to demonstrated entanglement fidelities of 0.991 \cite{madjarov2020} and projected gate fidelities of $\approx 0.999$ \cite{Saffman2016, saffman2019} for conventional Rydberg blockade gates. Importantly, this can be realized without ground state cooling of the atomic motion:  Doppler shifts are negligible for microwave transitions between Rydberg states, and the gates can be made insensitive to motion by exploiting the unique feature that the timescale of the atomic motion is comparable to the gate time and can be averaged out, to first order, with carefully chosen parameters. The proposed techniques may be implemented with a variety of atomic species, including alkali and alkaline-earth \cite{madjarov2020,Norcia2018,Saskin2019} atoms, and we discuss strategies applicable to both but give specific states and numbers relevant to rubidium.

\section{Overview}
\label{sec:overview}
We begin with an overview of the proposed scheme, before discussing the components in detail. Qubits are encoded in four circular Rydberg levels (Fig. \ref{fig:fig1}a): a pair of storage states $\{\ket{0_s},\ket{1_s}\} = \{\ket{59C},\ket{61C}\}$, and a pair of active states $\{\ket{0_a},\ket{1_a}\} = \{\ket{56C},\ket{64C}\}$ ($\ket{nC}$ denotes the circular state with $\ket{n,\ell,m} = \ket{n,n-1,n-1}$). The active states are used to realize two-qubit gates, while the storage states are ideally completely non-interacting. In practice, all Rydberg states interact with each other, so the implementation of effectively non-interacting storage states relies on tuning the interaction between them to precisely dipolar form, where it can be cancelled using global dynamical decoupling (DD) sequences such as WAHUHA \cite{Waugh1968}. The particular states selected are motivated by tuning the interactions into this form, as outlined in Section \ref{sec:gates}.

The circular atoms are stored inside an engineered microwave structure to realize long radiative lifetimes (Fig. \ref{fig:fig1}c; Section \ref{sec:cav}). They are individually confined in an array of ``compute" traps (in the $xy$-plane) with spacing $a_{circ} \approx 12\,\mu$m, which reflects a tradeoff between achieving strong interactions and staying in the perturbative regime of the van der Waals interaction (we focus here on a 1D array, but the extension to 2D is straightforward). The quantization axis is defined by parallel $E$ and $B$ fields perpendicular to the array, along $\hat{z}$. A second array of ancilla atoms is displaced along $\hat{z}$ by $d_z = 5\,\mu$m. The ancilla array allows the state of the circular atoms to be measured, by exploiting the Rydberg blockade between a low-$\ell$ Rydberg state of the ancilla atom, $\ket{a} = 55S$ (in Rb), and a circular atom in the compute array. By engineering a F\"orster resonance between the ancilla state and a particular circular state, this interaction can be made highly selective to a single circular state, chosen to be $\ket{1_a}$ (Section \ref{sec:meas}).

The experimental sequence proceeds as follows (Fig. \ref{fig:fig1}b). First, the compute and ancilla arrays are initialized with single ground-state atoms using rearrangement-based techniques \cite{Kim2016,Endres2016,Barredo2016}. Second, the compute array is excited into $\ket{1_a}$ using a combination of laser excitation and RF rapid adiabatic passage \cite{Nussenzveig1993}. Since this excitation is challenging to realize with extremely high fidelity (the state of the art is 90-95\% \cite{Signoles2017,Teixeira2020}), the ancilla array is used to nondestructively measure which atoms have been correctly excited, and a second rearrangement is performed to fill in a small number of defects in the circular atom array. Then, the atoms are transferred to $\ket{0_s}$, and the computation begins.

A sequence of gates is carried out by locally manipulating between circular states using the orbital angular momentum of focused LG beams (Section \ref{sec:manip}). Single-qubit gates are realized by driving the $\ket{0_s}-\ket{1_s}$ transition with $\Delta n = 2$. Two-qubit gates are realized by transferring a pair of atoms from the storage states to the active states using the operation $\Pi_{sa} = \ket{0_a}\bra{0_s} + \ket{1_a}\bra{1_s} + h.c.$ on each atom, with $\Delta n = 3$. 

Since the storage states are weakly interacting, a DD sequence with period $t_c$ is applied to these states with a global microwave drive, which removes the effect of interactions between them at certain refocusing times, $N t_c$. By executing single qubit gates at the refocusing times, and synchronizing the two-qubit gates with this cycle, the effect of the storage-storage and storage-active interactions is removed, leaving only a simple Ising-type interaction between the active states, which drives the two qubit gate (Section \ref{sec:gates}). Furthermore, for an appropriately designed DD sequence, errors arising from atomic motion are also suppressed if $t_c$ is commensurate with the motional period (Section \ref{sec:errors}).

At the end of the computation, the state $\ket{1_s}$ is transferred back to $\ket{1_a}$ for measurement using the ancilla array. While this can be done globally with microwaves to measure the entire array, a subset of the array can be measured by applying $\Pi_{sa}$ locally to a subset of sites. The atoms left behind in the storage states (with the dynamical decoupling sequence continuously applied) are unaffected by the measurement, allowing for partial readout of the qubit array.

\section{Engineering the local density of states}
\label{sec:cav}

\begin{figure}
    \centering
    \includegraphics[width=3.3 in]{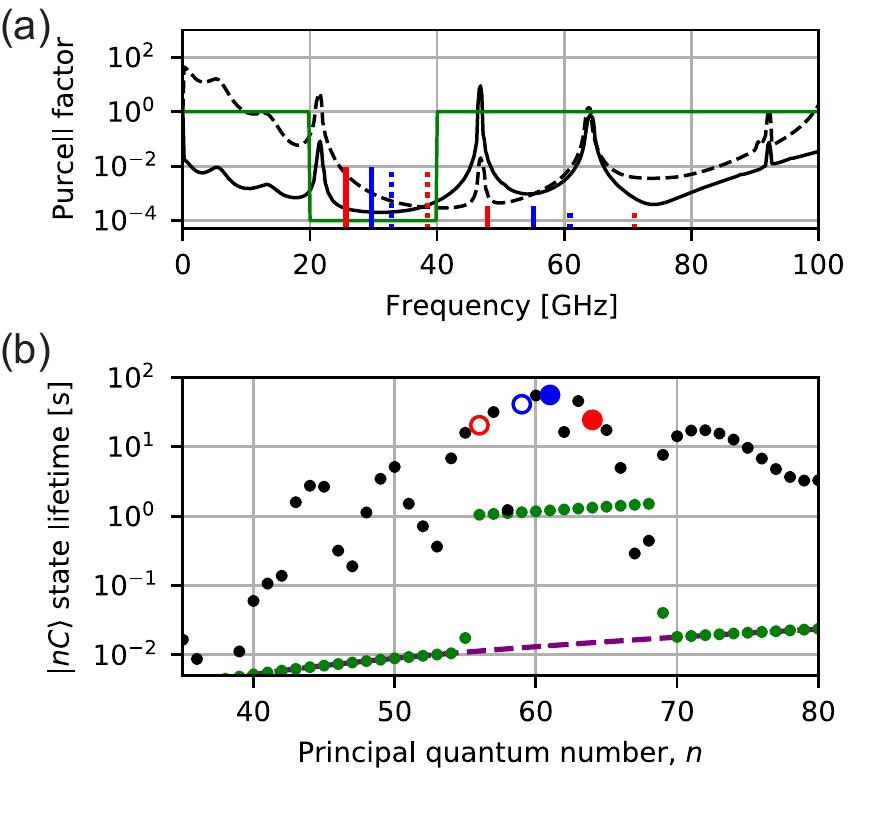}
    \caption{Extending circular state lifetimes. (a) LDOS suppression (Purcell) factor inside the microwave waveguide. The black solid (dashed) line shows the simulated $\sigma^{\pm}$ ($z$)-polarized LDOS at the center of the structure, while the green line shows a idealized model for comparison. Long (short) lines show the frequency of transitions between the $n$ and $n-1$ ($n+2$) manifolds for the four computational states. (b) Calculated circular Rydberg state lifetime for the LDOS shown in (a) at an environment temperature $T_b=4$ K. For the full structure (black), the states used in the protocol are indicated using the colors from Fig. \ref{fig:fig1}a. The dashed line shows the free space lifetime at the same temperature.}
    \label{fig:cavity}
\end{figure}

The heart of the proposed apparatus is a microwave structure that extends circular Rydberg atom lifetimes by suppressing the local density of states (LDOS) at frequencies where these states can absorb or emit radiation. The structure must also provide high-NA optical access for tweezers and single-atom imaging, and the ability to apply uniform, parallel electric and magnetic fields to define a quantization axis for the circular states.

Previous approaches to LDOS engineering with circular Rydberg atoms have used a parallel plate capacitor geometry \cite{Hulet1985,Nguyen2018}. This design suppresses the LDOS for in-plane electric field polarization (sufficient to extend the circular state lifetime) and allows the application of uniform electric fields. However, it does not provide high-NA optical access, and drilling a holes or inserting lenses into the capacitor breaks the translational symmetry, which mixes in- and out-of-plane fields and spoils the LDOS suppression. One possible workaround is thin, transparent oxide coatings such as indium tin oxide (ITO) \cite{Jau2015,meinertIndiumTinOxide2020}, although previous experiments have observed decreased Rydberg state lifetime near laser-illuminated ITO surfaces arising from an unknown mechanism \cite{Jau2015}.

On the other hand, a hollow, circular waveguide with diameter $D$ suppresses the LDOS for all polarizations below a cutoff frequency $f_c = 1.841 c/(\pi D)$ \cite{kleppner1981, pozarMicrowaveEngineering2011} and provides high-NA optical access through the end or using lenses inserted into the waveguide itself. However, such a structure also shields static electric fields very effectively, preventing the application of a bias field.

We propose a hybrid structure that consists of a pair of parallel, annular electrodes placed inside a waveguide (Fig. \ref{fig:fig1}c). A bias voltage is applied to the electrodes via a transmission line passing through the waveguide wall with an embedded, reflective low-pass filter to suppress microwave leakage. Choosing an appropriate electrode geometry allows fairly uniform electric fields in the waveguide center, with only quartic and higher dependence on the radial coordinate. Cross-bores of nearly the size of the central waveguide can be drilled in the side to allow lasers or atoms to pass through without affecting the microwave properties. Additional cross-bores can be used to apply microwaves with a controlled polarization via weakly coupled antennas.

The simulated LDOS of a representative structure is shown in Fig. \ref{fig:cavity}a. The spectrum crudely resembles a bandstop filter (Fig. \ref{fig:cavity}a), with the lower cutoff frequency determined by the electrode filter ($f_L \approx 3$ GHz) and the upper cutoff determined by the waveguide ($f_U \approx 80$ GHz). Within the stopband, the LDOS has a finite value $P_{min} \approx 10^{-4}$ because of ohmic losses in the gold walls, as well as several discrete resonances arising from the filter and electrode structure (additional details about the design and simulation can be found in Appendix \ref{app:waveguide}).

The calculated lifetime of circular states in this structure is shown in Fig. \ref{fig:cavity}b, assuming a blackbody temperature $T_b = 4$ K. The lifetime for the four computational states exceed 20 seconds, limited primarily by residual blackbody radiation. The lifetimes at $T_b=0$ K is approximately 5 times longer, limited purely by ohmic losses, while the lifetime at room temperature is approximately 200 ms, limited by blackbody radiation and increased ohmic losses associated with the higher resistivity of the metal walls. Based on additional considerations including photon scattering from optical traps, we estimate a useful lifetime of $\tau_{circ} \approx 3$ seconds (Appendix \ref{app:lifetime}).

To elucidate the role of the LDOS at different frequencies, we also plot the lifetimes for a simple bandstop model with $P = P_{min} = 10^{-4}$ from 20-40 GHz and $P=1$ elsewhere. At $T_b=0$ K, this would increase the lifetime of states from $n=56-68$ by $1/P_{min}$ from their free space values, to about 200 seconds. However, at $T_b=4$ K, the lifetime is limited to a few seconds by absorption of blackbody photons on transitions with $\Delta n = +2$. The full design extends the lifetime by 1-2 orders of magnitude more by suppressing the LDOS for these higher frequency transitions for both $\sigma^\pm$ and $\pi$ polarizations. Achieving the same suppression with a parallel plate capacitor that does not suppress $\pi$-polarized transitions would require freezing out the blackbody radiation with $T_b < 1$ K \cite{Nguyen2018}.

\section{State-insensitive optical traps}
\label{sec:trapping}

Leveraging the long circular state lifetimes requires that the atoms be trapped, ideally in a state-insensitive ``magic" trap to suppress motional decoherence. In this section, we discuss two approaches based on the ponderomotive potential (applicable to any atomic species), and the polarizability of the ion core (in alkaline earth atoms).

The nearly-free Rydberg electron experiences a ponderomotive energy shift in a laser field, which results in a center-of-mass potential for the Rydberg atom given by \cite{Dutta2000}:

\begin{equation}
\label{eq:up}
    U_n(\vec{R}) = \frac{e^2}{4 m_e \omega^2} \int |\psi_n(\vec{r})|^2 |\vec{E}(\vec{R}+\vec{r})|^2 d^3 \vec{r}
\end{equation}

where $e,m_e$ are the electron charge and mass, $\omega$ is the frequency of the optical field, $\vec{E}$ is the electric field of the laser and $\psi_n(\vec{r})$ is the wavefunction of the Rydberg electron in $\ket{nC}$ at position $\vec{r}$ with respect to the nucleus at $\vec{R}$. This effect has been exploited to confine Rydberg atoms in optical lattices \cite{Anderson2011}, hollow laser beams \cite{Cortinas2020} and focused bottle beams \cite{Barredo2020}.

However, a challenge with the ponderomotive potential is that it is inherently state-dependent, since the wavefunction $\psi_n$ varies between states. If the length scale over which the intensity $|E|^2$ varies is large compared to the extent of $\psi_n(\vec{r})$, then the integral in Eq. \eqref{eq:up} is independent of $n$ to some approximation. However, in optical tweezers used for single atom trapping and manipulation, the beam waist is typically below 1 $\mu$m, comparable to the extent of the Rydberg wavefunction ($\langle r \rangle = a_0 n^2 \approx 180$ nm for $\ket{59C}$). This gives rise to significant variation in the depth and shape of the potential across different Rydberg states \cite{Wilson2019}, and will result in significant motional decoherence.

\begin{figure}
    \centering
    \includegraphics[width=3.3 in]{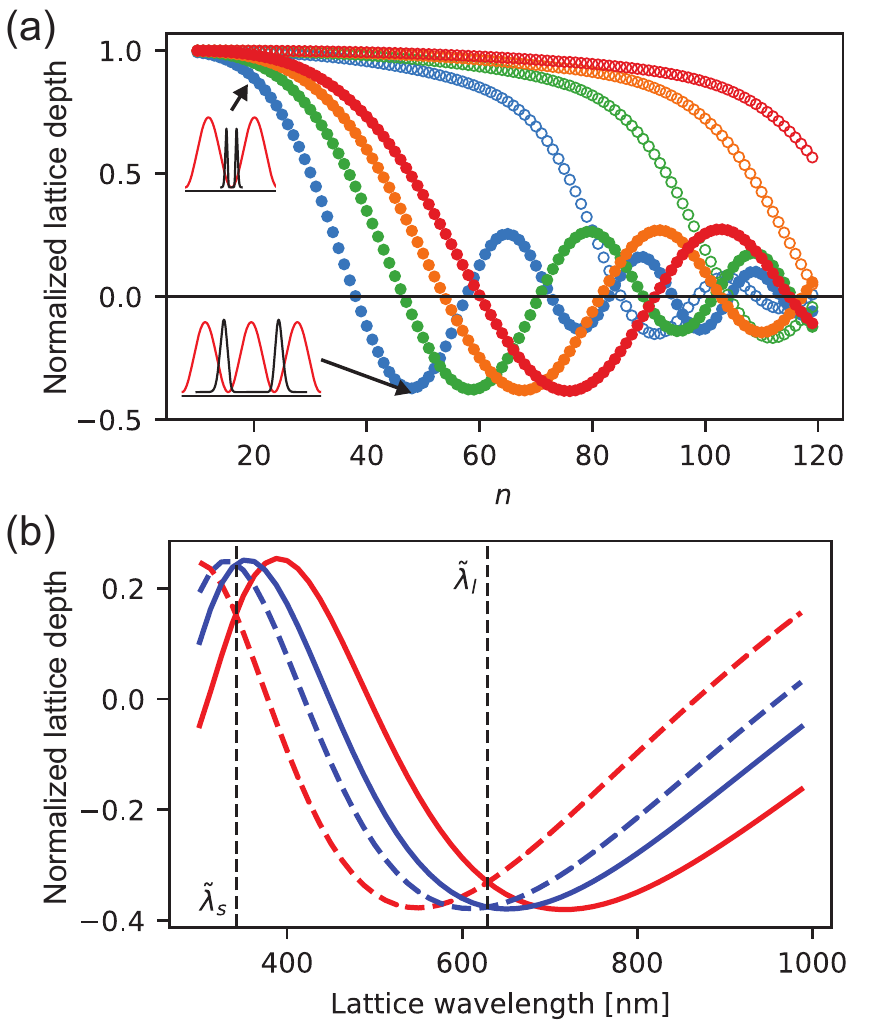}
    \caption{Ponderomotive optical lattices.
    (a) Normalized lattice depth [integral in Eq. \eqref{eq:pondlattice}] as a function of $n$, for an in-plane (filled circles) or $\hat{z}-$oriented (open circles) lattice. The colors denote different wavevectors $|k|=2\pi/\lambda$ with $\lambda = (400,600,800,1000)$ nm shown in (blue, green, orange, red). The insets show the configuration of the Rydberg wavefunction (black) with respect to the lattice (red) for the stable trap position at the indicated $n$.
    (b) Ponderomotive lattice depth for the four computational states as a function of wavelength, highlighting two wavelengths where the storage and active states have nearly identical trap depths.}
    \label{fig:lattice}
\end{figure}

However, in the special case of a lattice with $|E(\vec{R})|^2 = |E_0|^2 \left[1 + \cos(2\vec{k} \cdot \vec{R})\right]$, the ponderomotive potential can be written as:

\begin{equation}
\label{eq:pondlattice}
    \begin{split}
            U_n(\vec{R}) &= \frac{e^2 |E_0|^2}{4 m_e \omega^2} \times \\
            &\left[1+ \cos (2 \vec{k} \cdot \vec{R}) \int |\psi_n(\vec{r})|^2 \cos (2 \vec{k} \cdot \vec{r}) d^3 \vec{r}\right]
    \end{split}
\end{equation}

In this form, it is clear that the shape of the potential is the same for all states, and that the magic condition for a pair of states will be realized if the lattice depth given by the integral in Eq. \eqref{eq:pondlattice} is the same for both states. If the lattice vector $\vec{k}$ is perpendicular to the atomic quantization axis $\hat{z}$, the value of this integral is oscillatory in $n$ \cite{Dutta2000}, as the wavefunction expands to cover multiple periods of the lattice (Fig. \ref{fig:lattice}a). The presence of turning points in this non-monotonic dependence results in identical trapping potentials for pairs of Rydberg states with symmetric displacement from the turning points. This is sufficient to realize simultaneous magic traps for the pairs $\{\ket{0_s},\ket{1_s}\}$ and $\{\ket{0_a},\ket{1_a}\}$, which are symmetrically displaced around the same midpoint (Fig. \ref{fig:lattice}b).

To quantify this, we express the difference in trap potentials for the storage (active) states as $\eta_s$ ($\eta_a$), where $\eta = (U_1 - U_0)/U_0$ and $U_{0/1}$ refers to the depth for the state 0 or 1. It is not possible to make $\eta_a = \eta_s = 0$ exactly. However, exact magic wavelengths can be found for either pair near $\tilde{\lambda}_l$ (Fig. \ref{fig:lattice}b): $\lambda = 629.38$ nm gives $(\eta_a,\eta_s)=(0,6\times 10^{-4})$, and $\lambda=628.87$ nm gives $(\eta_a,\eta_s)=(3.5\times10^{-3},0)$. A compromise is also possible: $\lambda = 629.285$ nm gives $\eta_s \approx \eta_a \approx 6 \times 10^{-4}$. For comparison, $\eta_a \approx 0.2$ at $\lambda=800$ nm.

We note that there is a second magic wavelength for the in-plane lattice ($\tilde{\lambda}_s$ in Fig. \ref{fig:lattice}b). In addition to being at a less-convenient wavelength, the shorter period results in significant mixing between circular and elliptical states (\emph{i.e.}, nearly circular states, with $0 \ll |m| < n-1$). Near $\tilde{\lambda}_l$, this effect is negligible if the trap depth is no more than a few MHz.

Along the $\hat{z}$ direction, the behavior is quite different: the extent of the wavefunction is much smaller and it varies slowly with $n$. In this case, a nearly magic trap can be realized by choosing a $\hat{z}$-oriented lattice with sufficiently long period $\Lambda_z$. If $\Lambda_z = 800$ nm, $(\eta_a,\eta_s) = (4 \times 10^{-3}, 1 \times 10^{-3})$. Further reduction can be achieved by increasing $\Lambda_z$ (\emph{i.e.}, using a shallow intersection angle), with $\eta \propto \Lambda_z^{-2}$.

Realizing a 1 MHz deep lattice in the in-plane direction using a retro-reflected 629 nm beam requires an intensity $I_{XY} = 3 \times 10^6$ W/cm$^2$, corresponding to a one-way power of 11.8 W in a beam with a waist of $100\,\mu$m. This power is achievable with a buildup cavity \cite{heinzCrossedOpticalCavities2021}. In the out-of-plane direction, realizing the same lattice depth with a retro-reflected 1560 nm beam of the same size requires 0.7 W of one-way power.

In alkaline earth atoms, the dipole polarizability of the ion core makes an additional contribution to the trap potential \cite{Mukherjee2011,topcu2014}, as demonstrated for Yb \cite{Wilson2019} and Sr \cite{Teixeira2020}. For the in-plane 629 nm lattice, this has the effect of reducing the required intensity by a factor of approximately 3.5 for Sr (1.8 for Yb) because the polarizability of the ion core increases the lattice depth. Since the potential arising from the ion core is completely independent of the state of the circular electron, the value of $\eta$ is reduced by the same factor. For the vertical 1560 nm lattice, the ion core polarizability decreases the lattice depth, but only by a few percent.

This suggests another approach in alkaline earth atoms, which is to trap at a wavelength very close to an ion core resonance, such that the contribution from the ion core polarizability is much larger than the ponderomotive potential. In that case, it may be possible to use a red-detuned, Gaussian optical tweezer, even though the variation in the ponderomotive potential is quite large between states \cite{Wilson2019}. At a detuning of $\Delta_t \approx - 1.5$ THz from the $^2S_{1/2} \rightarrow ^2P_{1/2}$ transition in either Yb$^+$ (369 nm) or Sr$^+$ (422 nm), the AC Stark shift from the ion core is approximately 100 times greater than the ponderomotive potential, such that $\eta_a,\eta_s \lesssim 10^{-3}$ can be realized in a tweezer with beam waist $w_0 = 500$ nm. This will result in a photon scattering rate of approximately $50-100\,\,\textrm{s}^{-1}$ for a 1 MHz deep trap, which is significant but potentially manageable in view of the fact that these events should not disturb the state of the circular electron. In fact, the ion core transitions may even be used for continuous laser cooling \cite{Teixeira2020}. The inconvenience of using UV or blue wavelengths may be offset by the small powers required at such close detunings: less than 100 $\mu$W per tweezer is sufficient to realize a 1 MHz trap depth.

Details about the lifetime and loading procedure for the traps are discussed in appendices \ref{app:lifetime} and \ref{app:loading}, respectively.

\section{Nondestructive measurement of circular Rydberg atoms} \label{sec:meas}

\begin{figure}
    \centering
    \includegraphics[width=3 in]{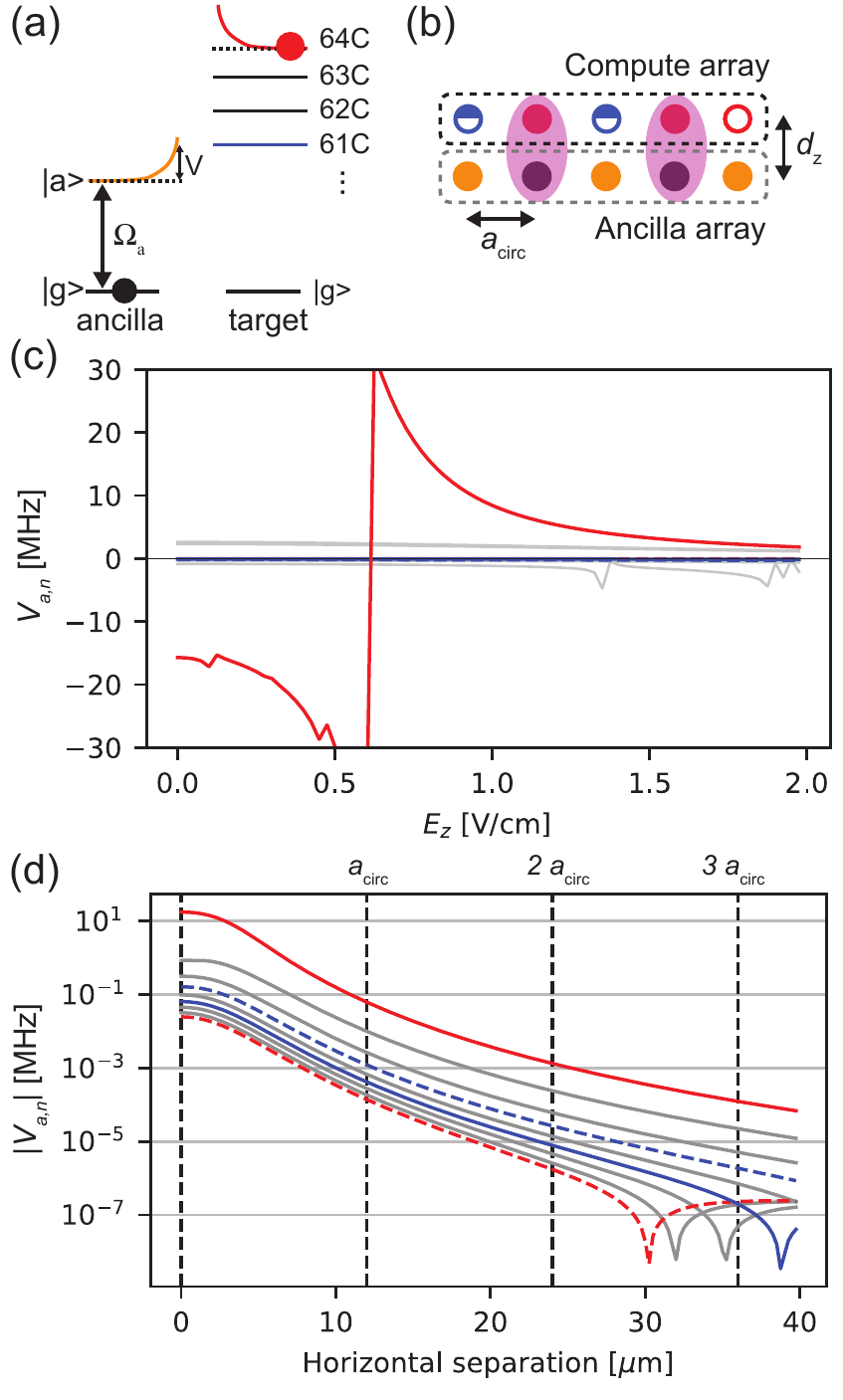}
    \caption{Nondestructive circular state detection.
    (a) The measurement is realized by a state-selective F\"orster resonance between a Rb ancilla in $\ket{a} = 55S$ and $\ket{64C}$.
    (b) An array of ancilla atoms displaced by $d_z = 5\,\mu$m are used to probe the state of the circular atom array. The excitation to $\ket{a}$ is blockaded on sites with a compute atom in $\ket{1_a}$. 
    (c) Van der Waals interaction $V_{a,n}$ between $\ket{a}$ and $\ket{nC}$, as a function of electric field $E_z$.  The states $n=56-66$ are shown, with the non-computational states in gray.
    (d) $|V_{a,n}|$ at $E_z=0.25$ V/cm, with varying lateral separation.
    }
    \label{fig:detect}
\end{figure}

Another crucial ingredient of the proposed architecture is a method to non-destructively detect the states of circular Rydberg atoms. Previous work with circular Rydberg atoms has relied on measurements via state-selective field ionization \cite{Hulet1983,Gleyzes2007,Teixeira2020}, which are inherently destructive to the atoms. In this work, we focus on a novel technique that is rapid, site-resolved and non-destructive (QND). This approach is based on the van der Waals interaction between the circular atom to be measured and a low-$\ell$ state, $\ket{a}$, of a nearby ancilla atom (Fig. \ref{fig:detect}a). Specifically, by exploiting a F\"orster resonance between one of the computational states ($\ket{1_a}$) and $\ket{a}$, the magnitude of van der Waals interaction with $\ket{1_a}$ can be made significantly larger than for the other computational states. Probing the Rydberg blockade of the ancilla atom then yields information about the state of the circular atom.

In Fig. \ref{fig:detect}c, we show the interaction $V_{a,n}$ between a Rb ancilla in $\ket{a} = 55S_{1/2}, m_J=-1/2$, and a target atom in various circular states from $n=56$ to $n=66$. When the electric field $E_z=0$ V/cm, $V_{a,64}$ is two orders of magnitude larger than any of the other three computational states, and this ratio can be made even larger by approaching an exact F\"orster resonance around $E_z = 0.6$ V/cm. We note that the interaction is essentially the same for elliptical states of the same $n$ (not shown), so the measurement is better described as selective for the $n=64$ manifold instead of the single state $\ket{1_a}$. Similar resonances exist for any atomic species.

This selective interaction can be used to probe the state of a circular Rydberg atom using a standard Rydberg blockade controlled-$Z$ (CZ) gate on the ancilla atom \cite{Lukin2001}. Given an ancilla prepared in a superposition of two ground states, $\ket{\psi} = (\ket{g}+\ket{g'})/\sqrt{2}$, a $2\pi$ pulse on the ancilla $\ket{g}$ to $\ket{a}$ transition with Rabi frequency $\Omega_a$ imparts a $\pi$ phase shift on $\ket{g}$ only if it is not blockaded by the target atom. The resulting spin rotation can be measured using a subsequent $\pi/2$ rotation on the ancilla and fluorescence detection distinguishing $\ket{g}$ and $\ket{g'}$. Errors can arise from the finite blockade strength as well as the finite lifetime of the ancilla state, and are minimized at an optimum value of the Rabi frequency $\tilde{\Omega}_a = (\pi V_{a,64}^2/\tau_a)^{1/3} \approx 2 \pi \times 1$ MHz (here, we take $V_{a,64} = 2\pi \times 20$ MHz and an ancilla lifetime $\tau_a = 200\,\mu$s, appropriate for cryogenic temperatures). The resulting error probability is $P_g = \left[\pi/(V_{a,64} \tau_a)\right]^{2/3}/2 \approx 1.3 \times 10^{-3}$, assuming perfect initialization and readout of the ancilla spin state. This is the same error scaling as a conventional Rydberg blockade gate \cite{Saffman2016}, but with a smaller prefactor since the circular atom has negligible decay over the gate. Because of its QND nature, the measurement can be repeated to improve the accuracy \cite{hume2007}.

Next, we consider cross-talk between adjacent qubits. Fig. \ref{fig:detect}d shows the interaction strength between an ancilla and neighboring circular Rydberg atoms in the compute array. The interaction decays rapidly with distance, and for $a_{circ}=12\,\mu$m, the interaction with off-target sites is suppressed by a factor of 300, ensuring highly site-resolved measurements.

To measure the entire array, it is sufficient to transfer the population from $\ket{1_s}$ to $\ket{1_a}$ on every site before attempting to excite the ancillae. It is also possible to measure a subset of the array, by transferring a subset of the atoms from the storage to the active states using $\Pi_{sa} = \ket{0_a}\bra{0_s} + \ket{1_a}\bra{1_s} + h.c.$. The atoms that are not to be measured are left in the storage states, and as long as their ancilla atoms are not excited to $\ket{a}$ (\emph{i.e.}, by selective addressing with $\Omega_{a}$), they will be largely unaffected by the measurement of their neighbors, with induced errors below $10^{-6}$ (Appendix \ref{app:meas}).

The size of the array that can be initialized is limited by the measurement error as well as the decay of the circular atoms over the measurement and rearrangement time. Given the measurement error above and assuming the ancilla readout and rearrangement can be performed in approximately 10 ms for a decay probability of $P_d \approx 3 \times 10^{-3}$, arrays of $N \approx 250$ circular atoms can be realized with an average of one defect.

\section{Ponderomotive manipulation of circular Rydberg states} 
\label{sec:manip}

Another key component of the proposed scheme is a mechanism to locally manipulate atoms between circular states. In current experiments with circular Rydberg states, transitions between circular levels are driven using microwave electric fields coupled to the strong electric dipole transition \cite{Signoles2014,Signoles2017,Teixeira2020}. While providing extremely large Rabi frequencies, microwaves cannot be locally addressed, and are limited to transitions between states with $|\Delta m| = 1$ (in a single-photon transition).

Here, we discuss an alternative approach to manipulating the Rydberg electron, using the ponderomotive potential from focused LG beams. This enables locally-addressed operations, and the use of orbital angular momentum instead of the spin angular momentum of the photon allows driving transitions with $|\Delta m| = 2$ or $3$ in a single step, covering all of the transitions between computational states indicated in Fig. \ref{fig:fig1}a.

The ability of the ponderomotive potential to couple states of different angular momentum has been recognized since the earliest proposal for ponderomotive traps for Rydberg atoms \cite{Dutta2000,Knuffman2007}, and coupling of degenerate levels by a static intensity (essentially a high-rank tensor light shift) has been experimentally demonstrated \cite{Anderson2012,Wilson2019}. Superimposing very high-order LG beams with different frequencies to drive direct transitions from low-$\ell$ to circular Rydberg states was recently proposed in Ref. \cite{Cardman2020}. Here, we extend this idea in a few ways. First, we apply the concept to $|\Delta m| = 2$ or 3 transitions between nearby circular states, and observe that it can be quite efficient because of the good overlap between these LG modes and the circular wavefunctions. Second, we estimate the infidelity associated with photon scattering, and conclude that the error per gate is comparable to hyperfine spin manipulation in atomic ground states using two-photon Raman transitions.

To compute the strength of these transitions, we recast the ponderomotive potential in Eq. \eqref{eq:up} as a matrix element between two states:

\begin{equation}
    \label{eq:pondme}
    \bra{\psi'}U_p\ket{\psi} = \frac{e^2}{4 m_e \omega^2} \int \psi'^*(\vec{r}) |\vec{E}(\vec{R}+\vec{r})|^2 \psi(\vec{r}) d^3 \vec{r}
\end{equation}

The incident electric field is taken to be the sum of two co-propagating LG modes. In the paraxial limit, these are described in cylindrical coordinates by a function of the form $\vec{E}_{lm}(r,z,\phi) = E_{lm}(r,z)e^{i m \phi} \hat{x}$, where $\hat{x}$ is the polarization direction and $\hat{z}$ is the propagation direction, parallel to the quantization axis of the atomic states \cite{siegman1986}. The total field intensity becomes:

\begin{align}
\begin{split}
    &|E|^2 = |E_{lm}(r,z)|^2 + |E_{l'm'}(r,z)|^2 \\
    & + 2 \Re \left[E_{lm}(r,z)E^*_{l'm'}(r,z) e^{- i (\omega-\omega') t - i (m-m') \phi} \right]
\end{split}
\end{align}

where $(l,m,\omega)$ are the mode numbers and optical frequency for the first beam, and the primed quantities are for the second beam.

The interference term generates a rotating intensity pattern $|E|^2 \propto \cos\left[(\omega-\omega') t - (m-m') \phi\right]$ which can resonantly drive transitions between atomic states separated by angular momentum $\Delta m = m-m'$ and energy $\Delta E = \hbar(\omega-\omega')$. The matrix element can be evaluated by integrating Eq. \eqref{eq:pondme} directly or expanding the operator $|\vec{E}(\vec{R}+\vec{r})|^2$ in the basis of spherical harmonics \cite{Wilson2019}. The localized nature of the circular wavefunctions also enables an accurate approximation of the Rabi frequency $\Omega_{n',n} = \mel{n'C}{U_p}{nC}/\hbar$ using $\psi_{n'}^*\psi_n = \delta(r-r_0)\delta(z) e^{i(n-n')\phi}/(2\pi r_0)$ with the average radius $r_0 = \frac{1}{2}a_0 (n^2 + n'^2)$. If the beam is centered on the atomic nucleus (\emph{i.e.}, $\vec{R}=0$), then the transition matrix element in the frame rotating at $\omega-\omega'$ is:

\begin{equation}
\label{eq:pondapprox}
    \Omega_{n',n} = \frac{e^2}{4 \hbar m_e \omega^2} \Re\left[E_{lm}(r_0,0)E^*_{l'm'}(r_0,0)\right] \delta_{\Delta n,\Delta m}
\end{equation}

The order of magnitude of $\Omega_{n,n'}$ can be estimated by considering the ponderomotive shift arising from a single, Gaussian beam ($E_{00}$) in the limit of small $n$ where $E(r_0,0) \approx E(0,0)$. For a beam focused to a waist $w_0=\lambda$, $U_p/(\hbar P) = e^2/(4 \pi^3 \hbar \epsilon_0 c^3 m_e) = 2 \pi \times 1.44$ MHz/mW where $P$ is the optical power and $\epsilon_0$ is the permittivity of free space. In this expression, the $1/\omega^2$ dependence of the ponderomotive polarizability cancels with the increased focusing at shorter wavelengths, illustrating that for a fixed numerical aperture, the optimum wavelength is where the spatial overlap with the Rydberg electron wavefunction is maximized.

\begin{figure}
    \centering
    \includegraphics[width=3 in]{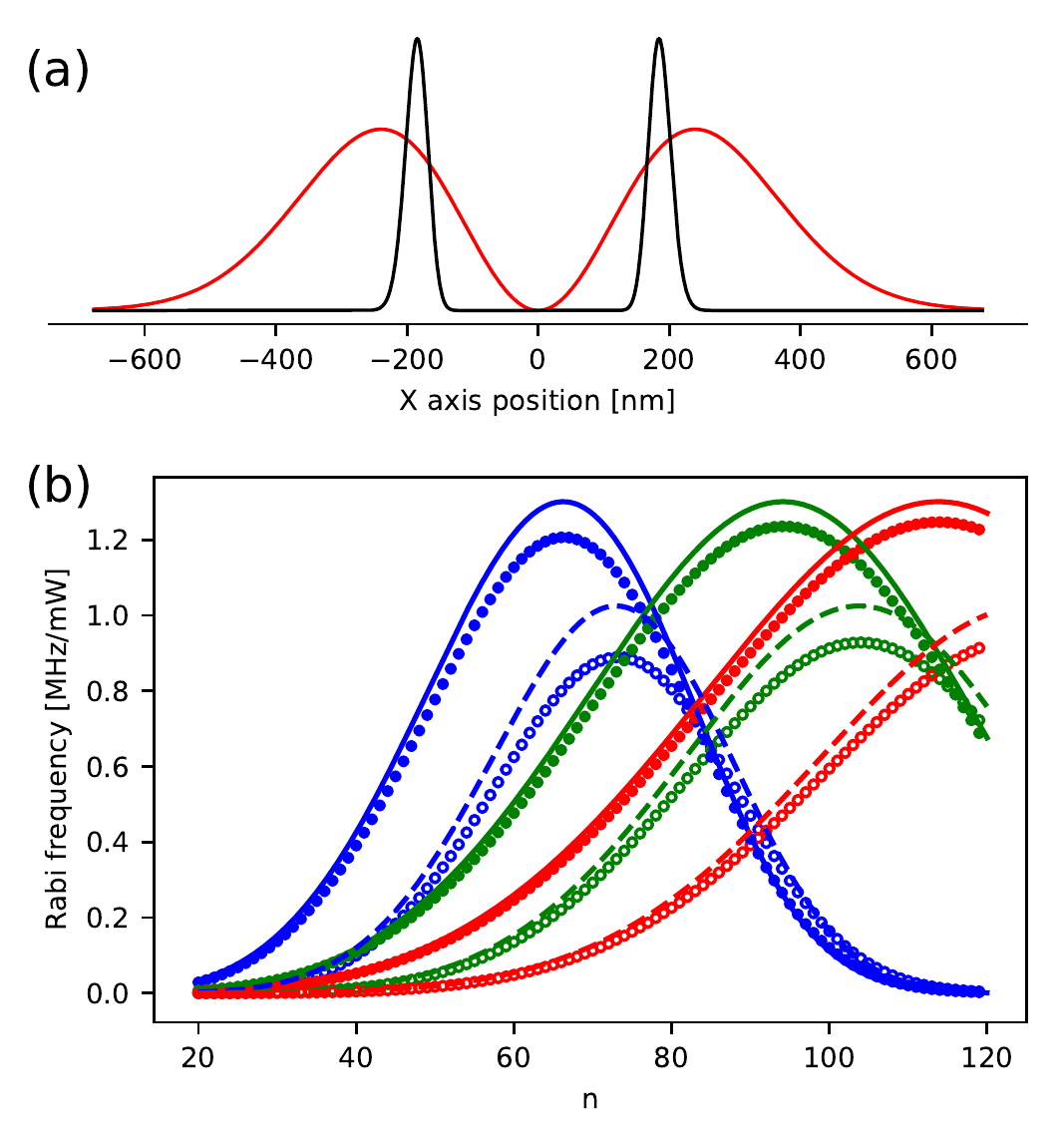}
    \caption{Local manipulation using LG modes. (a) Overlap of the radial $\ket{59C}$ wavefunction (black) with $LG_{0,1}$ (red, $\lambda=532$ nm, NA=0.5). (b) $\Omega_{n,n'}$ for different wavelengths focused with NA=0.5: 532 nm (blue), 1064 nm (green), and 1550 nm (red). Filled symbols show $n' = n+2$ transitions driven by interfering $LG_{0,-1}$ and $LG_{0,1}$ modes, while open symbols show $n' = n+3$ driven by interfering $LG_{0,-1}$ and $LG_{0,2}$ modes. The corresponding lines show the approximation Eq. \eqref{eq:pondapprox}.}
    \label{fig:driving}
\end{figure}

In Fig. \ref{fig:driving}, we estimate relevant parameters for the states used in the experiment. Fig. \ref{fig:driving}a shows that states near $\ket{60C}$ have good overlap with LG$_{0,1}$ at $\lambda=532$ nm (focused with NA=0.5). In Fig. 5b, we show the computed Rabi frequencies for the combinations LG$_{0,-1}$ + LG$_{0,1}$ (driving $\Delta m =2$) and LG$_{0,-1}$ +LG$_{0,2}$ ($\Delta m=3$), for several wavelengths. In both cases, $\lambda = 532$ nm is nearly optimal, but the penalty for using longer wavelengths is not excessive and confers the additional benefit of being less sensitive to variation in $\vec{R}$ (see section \ref{sec:errors}). The maximum Rabi frequencies are approximately 1 MHz per mW of power in each of the two beams, such that achieving drive strengths in excess of 10 MHz is practical.

Lastly, we consider incoherent errors, specifically, Thomson scattering of photons by the nearly-free Rydberg electron \cite{Nguyen2018}. Physically, this corresponds to radiation from the dipole formed by the electron oscillating in the laser field \cite{crowley2014}. The total scattering rate is given by $\Gamma = I \sigma_T/(\hbar \omega)$, where $\sigma_T = (8 \pi/3)(e^2/4 \pi \epsilon_0 m c^2)^2$ is the Thomson scattering cross section of the electron and $I=\epsilon_0 c |E|^2/2$ is the light intensity. Given a matrix element $U_p$, the error probability per $\pi$ pulse is $\epsilon_{\pi} = \frac{I \sigma_T/(\hbar \omega)}{U_p/(2h)} = 4 e^2 \omega / (3 \epsilon_0 m_e c^2) \approx 1 \times 10^{-6}$. We note that this compares favorably to the fidelity of a Raman transition between two spin ground states in Rb at the optimal detuning, which is $\epsilon_{\pi} = 2 \sqrt{2} \pi \Gamma/\Delta_{FS} \approx 8 \times 10^{-6}$ [$\Gamma/(2\pi) = 6$ MHz is the transition linewidth, and $\Delta_{FS} = 7$ THz is the fine structure splitting] \cite{Ozeri2007}. However, 5-10 times more intensity is required to drive the ponderomotive transition than the Raman process at its optimal detuning, and there is no way to trade fidelity for intensity by moving closer to resonance.

\section{Gate implementation}
\label{sec:gates}
The final key result is a scheme for implementing arbitrary quantum logic operations using qubits encoded in circular levels. Unlike conventional Rydberg blockade gates where atoms are only excited into Rydberg states during a multi-qubit gate operation, here, the atoms are always in Rydberg states and therefore always interacting to some degree. Managing unwanted interactions is the central challenge. We address this using four circular Rydberg levels: a pair of ``active" levels $\ket{0_a}, \ket{1_a}$ for executing two-qubit gates, and a pair of ``storage" levels $\ket{0_s}, \ket{1_s}$ for storing qubits and  implementing single-qubit gates.

These levels are selected so the interaction Hamiltonian $H_{int} = H_{ss} + H_{sa} + H_{aa}$ takes the form:

\begin{align}
    \label{eq:Hss}
    H_{ss} &= \sum_{ij} J^{ss}_{z} S^i_z S^j_z + J^{ss} \left(S^i_x S^j_x + S^i_y S^j_y\right) + \Delta^{ss} S_z^i n^j \\
    \label{eq:Hsa}
    H_{sa} &= \sum_{ij} J^{sa}_z S^i_z \bar{S}^j_z + \Delta^{sa} S_z^i \bar{n}^j + \Delta^{as} \bar{S}_z^j n^i\\ 
    \label{eq:Haa}
    H_{aa} &= \sum_{ij} J^{aa}_z  \bar{S}^i_z \bar{S}^j_z + \Delta^{aa} \bar{S}^i_z \bar{n}^j
\end{align}

Here, $S$ and $\bar{S}$ are pseudo spin-1/2 operators acting on the subspaces $\ket{0_s}$, $\ket{1_s}$, and $\ket{0_a}$, $\ket{1_a}$, respectively, while $n$ and $\bar{n}$ count the number of atoms on a site (0 or 1) in each subspace. The coefficients $J_z$ and $J$ give the strength of Ising and exchange interactions, respectively, while the $\Delta$ terms represent an effective local field depending on the presence (but not the state) of another atom. Each interaction coefficient has an implied dependence on the vector $\vec{R}_{ij}$ connecting the sites $i,j$.

Importantly, $H_{ss}$ contains both exchange and Ising terms, while $H_{sa}$ and $H_{aa}$ have only Ising couplings. This form of the Hamiltonian is guaranteed by the choices of the storage and active states. The storage states $\{\ket{0_s},\ket{1_s}\} = \{\ket{59C},\ket{61C}\}$ have $\Delta n=2$, such that the leading order Ising and exchange terms are both van der Waals, with dominant $1/R^6$ distance dependence. At the same time, the active states $\{\ket{0_a},\ket{1_a}\} = \{\ket{56C},\ket{64C}\}$ are separated by $\Delta n > 3$ from each other and from the storage states, such that the leading-order exchange terms are higher-order than quadrupole-quadrupole and can be neglected. Details about the computation of the interaction strengths can be found in Appendix \ref{app:interactions}.

\subsection{Storage states}

\begin{figure}
    \includegraphics[width=3.3in]{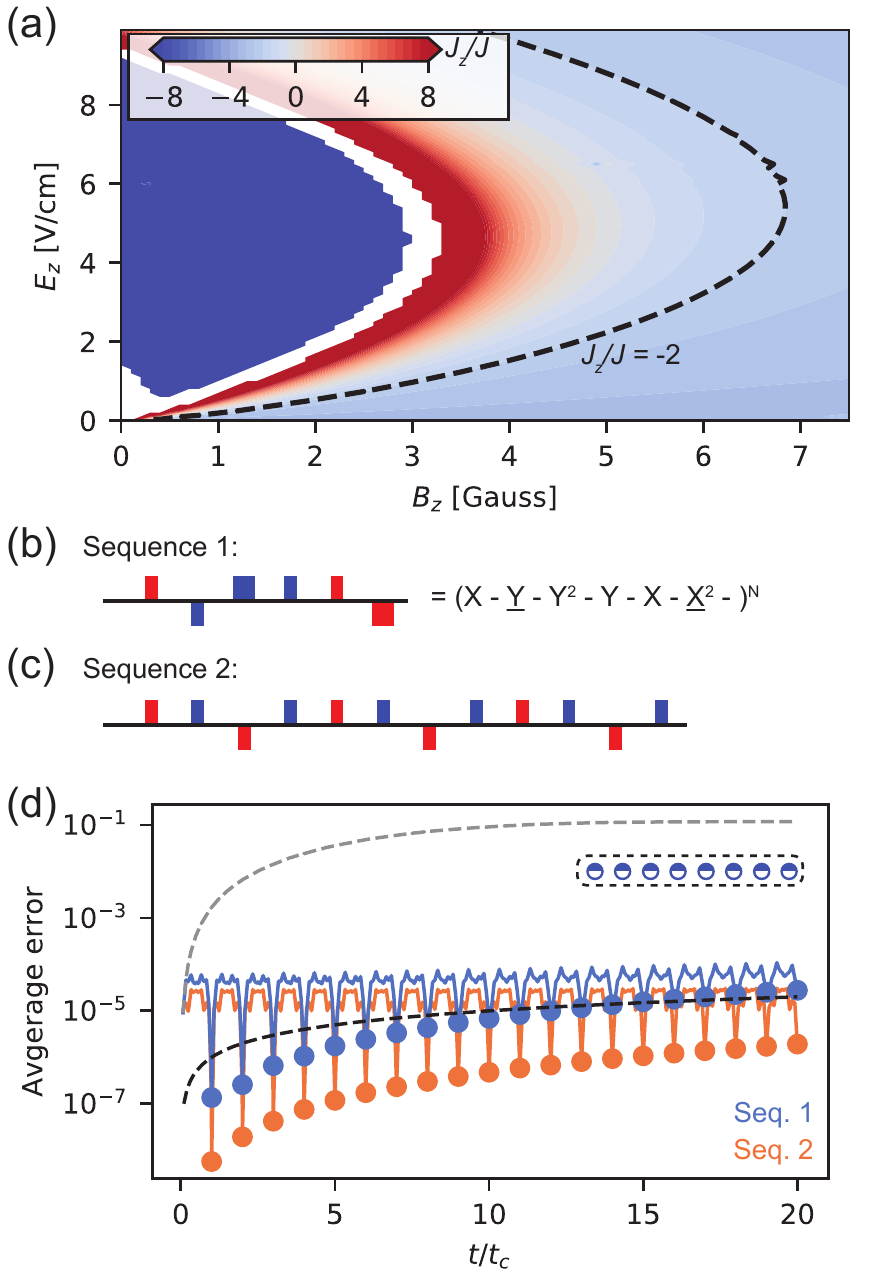}
    \caption{Decoupling storage state interactions. (a) $J_z^{ss}/J^{ss}$ as a function of electric and magnetic fields. The dashed line shows the dipolar condition $J^{ss}_z=-2J^{ss}$.
    (b) Sequence 1: 6-pulse WAHUHA-like sequence that cancels dipolar interactions and disorder. Red (blue) pulses denote $\pm \pi/2$ rotations around the $\hat{x}$ ($\hat{y}$) axis, with the minus sign for pulses below the line. The longer pulses are $\pi$ rotations. In the text representation, $X,Y$ are $\pi/2$ rotations ($-\pi/2$ if underlined).
    (c) Sequence 2: 12-pulse sequence with better robustness to finite pulse duration, pulse angle errors, and thermal motion (Section \ref{sec:errors}).
    (d) Storage error for an 8-atom chain without DD (grey dashed line), and with the sequences 1 (blue) and 2 (orange). For both, the sequence period is $t_c = 0.021/J^{ss} \approx 3.7\,\mu$s and the pulse duty cycle is $N_p t_p/t_c = 2.5 \times 10^{-2}$. The enlarged data points show the fidelity at the refocusing times $t=N t_c$. The black dashed line shows the incoherent error resulting from a finite circular lifetime $\tau_{circ}=3$ s, for comparison.
    \label{fig:int_storage}}
\end{figure}

Let us first consider the behavior of an array of atoms in the storage states. When the condition $J^{ss}_{z} = -2 J^{ss}$ is satisfied, Eq. \eqref{eq:Hss} is the dipolar interaction Hamiltonian. The influence of $H_{ss}$ (with $\Delta^{ss}=0$) on all pairs of atoms can be cancelled using \emph{global} dynamical decoupling (DD) sequences that symmetrize the interaction, such as the famous WAHUHA sequence developed for solid-state NMR \cite{Waugh1968}. Under repeated application of such a sequence with period $t_c$, the many-body system returns to its initial state at times $N t_c$ (\emph{i.e.}, $U(N t_c) \approx \hat{I}$), to at least first order in $t_c$. By only driving gate operations or performing measurements at times $N t_c$, the effect of interactions in the storage state is eliminated. We note that while any interaction Hamiltonian can be cancelled using \emph{local} pulses, the resulting pulse sequences are long and complex as nearest neighbor interactions, next-nearest neighbor interactions, etc. must be separately decoupled in nested cycles \cite{Vandersypen2005}. Therefore, tuning $H_{ss}$ to the dipolar form not only reduces the need for local drives, but also reduces the sequence complexity considerably.

The dipolar condition $J^{ss}_{z} = -2 J^{ss}$ can be realized by tuning the electric and magnetic fields, as shown in Fig. \ref{fig:int_storage}a (the broad tunability of these interactions was previously discussed in Ref. \cite{Nguyen2018}). Changing the fields mainly affects $J^{ss}_{z}$. At a separation $a_{circ}=12\,\mu$m, $J^{ss} = 2 \pi \times 918$ Hz.

To probe the effectiveness of dynamical decoupling at preserving arbitrary many-body states, we numerically compute the evolution of an 8-atom chain in the storage states (Fig. \ref{fig:int_storage}d). We compare two sequences: sequence 1, a six-pulse WAHUHA-like sequence (Fig. \ref{fig:int_storage}b), and sequence 2, a 12-pulse sequence slightly modified from ``sequence G" in Ref. \cite{Choi2020} (Fig. \ref{fig:int_storage}c). These sequences are designed to cancel local disorder from $\Delta_{ss}$ in addition to the dipolar interactions, which makes them distinct from the original WAHUHA sequence that preserves local fields. We compare the propagator $U(t)$ of the full system to the propagator $U_0(t)$ resulting from the same pulse sequence but with $H_{ss} = 0$, and compute the average error per atom $\bar{\epsilon}_S = (1-|\bra{\psi}U_0^\dag(t) U(t) \ket{\psi}|^2)/N_a$ ($N_a=8$ is the number of atoms, and the fidelity is averaged over $\ket{\psi}$ sampled from Haar random states within the space $\{\ket{0_s},\ket{1_s}\}^{\otimes N_a}$). We consider a sequence period $t_c = 0.021/J^{ss} \approx 3.7\,\mu$s, chosen to match the two-qubit gates described in the following section, and a finite rotation strength corresponding to a pulse duty cycle $N_pt_p/t_c = 0.025$ (here, $N_p$ is the number of pulses in a cycle and $t_p$ is the pulse duration).

As seen in Fig. \ref{fig:int_storage}d, the error grows quickly without any DD, but both sequences refocus the state at times $N t_c$. After one cycle, the coherent error arising from imperfect refocusing is several orders of magnitude smaller than the incoherent error from the finite lifetime of the circular states ($t_c/\tau_{circ} \approx 1 \times 10^{-6}$ for these parameters). While the basic sequence 1 decouples the interactions very well, sequence 2 performs better and is also more robust to errors and thermal motion as discussed in Section \ref{sec:errors}. The most important parameter, however, is the pulse period: the error for both sequences scales as approximately $(t_c J^{ss})^4$ (Fig. \ref{fig:errors}a).

The coherent errors from residual interactions grow quadratically with time, and without mitigation will eventually exceed the incoherent errors that grow linearly, as seen for sequence 1 in Fig. \ref{fig:int_storage}d. The quadratic growth rate can be suppressed using longer sequences that cancel interactions to higher order in $t_c$, as is common in NMR \cite{Burum1979}. An alternative approach from the field of quantum computing is randomized compiling \cite{Wallman2016}, where random single-qubit \emph{twirling operators} are inserted to frustrate the coherent evolution of unitary errors. We have observed in numerical simulations that the insertion of random single-qubit rotations after each period of the DD sequence results in linear error growth. Since the form of the coherent errors is known analytically, even greater suppression may be possible with deterministic compilation of twirling operators in a particular circuit.

The DD sequence on the storage atoms can be driven using $\Delta m = 2$ ponderomotive transitions as described in Section \ref{sec:manip}. However, since the same pulses are applied to all atoms, it is preferable to use global microwave driving (via a two-photon transition). In the latter case, a small exchange interaction is introduced during the pulse from the population of the intermediate state $\ket{60C}$. This must be incorporated into the design of the DD sequence by tuning $H_{ss}$ slightly away from dipolar form.


\subsection{Two-qubit gates in the active states}

\begin{figure}
    \includegraphics[width=3.3in]{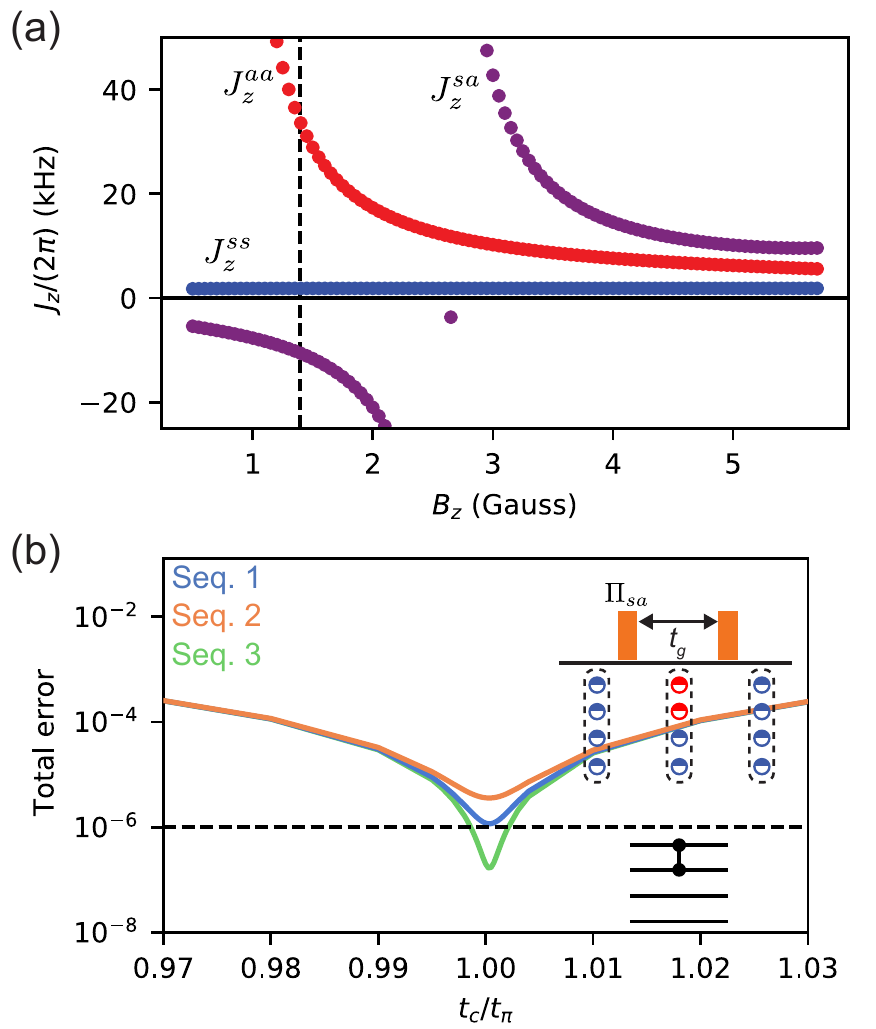}
    \caption{(a) Strength of the $J_z$ terms in Eqs. \eqref{eq:Hss}-\eqref{eq:Haa} as a function of $B_z$, moving along the lower branch of the $J_z^{ss}/J^{ss}=-2$ contour in Fig. \ref{fig:int_storage}(a). $J_z^{aa}$ and $J_z^{sa}$ both pass through F\"orster resonances, enabling wide control of their magnitude. At $B_z=1.39$ G (dashed line), $J_z^{aa}=2 \pi \times 33.6$ kHz, corresponding to $t_\pi=3.7\,\mu$s.
    (b) Error in a four-qubit array during a two-qubit gate, as a function of the gate duration $t_g$. The blue and orange curves correspond to sequences 1 and 2, while green is sequence 3 (see text and Fig. \ref{fig:sequences_SI}c).}
    \label{fig:int_active}
\end{figure}

To implement a two-qubit gate, a pair of atoms are moved from the storage subspace to the active subspace at a refocusing time $N t_c$, and returned at a later time $N' t_c$. The operation $\Pi_{sa} = \ket{0_a}\bra{0_s} + \ket{1_a}\bra{1_s} + h.c.$ is applied immediately before or after the last pulse in the DD sequence. While the atoms are in the active states, they are unaffected by the DD pulses on the storage states, which are far detuned from transitions affecting the active states. Therefore, the atoms interact under $H_{aa}$ for a time $t_g=(N'-N)t_c$, which will realize a controlled-Z (CZ) gate when $t_g = t_\pi = \pi/(4 J^{aa}_z)$. This is accompanied by a single-qubit rotation on each qubit, which can be compensated by adjusting the phase of one of the terms in $\Pi_{sa}$.

During the gate, atoms in the active states interact with spectator storage atoms through the $J_z^{sa}$ term in $H_{sa}$. However, since $\bar{S}_z$ is constant during the gate (it commutes with $H_{sa}$ and $H_{aa}$), it appears in $H_{sa}$ as a constant detuning for the storage atoms, which will be removed by the DD sequence along with the $\Delta$ terms. By the same token, the action of the spectator storage atoms on the active qubits is removed. It is crucial that $H_{sa}$ and $H_{aa}$ not have any exchange terms, as these would not be refocused in the same way.

The gate fidelity is fundamentally limited by the duration $t_\pi$, which determines the incoherent error probability over the gate cycle $\epsilon_\tau = t_\pi / \tau_{circ}$. The value of $J^{aa}_z$ can be tuned by adjusting the value of the $E$ and $B$ fields, as shown in Fig. \ref{fig:int_active}a. A F\"orster resonance at low fields gives wide tunability, and at the indicated field $B_z \approx 1.39$ G, a value of $t_\pi = 3.7\,\mu$s can be obtained with $J_z^{aa}/J_z^{ss} \approx 18$, resulting in a lifetime-limited error probability $\epsilon_\tau \approx 10^{-6}$.

In Fig. \ref{fig:int_active}b, we show the coherent error rate of a two-qubit gate in a four-atom array with two spectator qubits. The error rate is computed as  $\epsilon_{CZ} = 1-|\bra{\psi}U^\dag_{CZ}U\ket{\psi}|^2$. Here, $U$ is the numerically computed propagator and $U_{CZ}$ describes the ideal CZ gate (including single-qubit phases from $\Delta^{sa}$ and $\Delta^{aa}$), and an average is taken over Haar-random $\ket{\psi}$ from $\{\ket{0_s},\ket{1_s}\}^{\otimes 4}$ (the operation $\Pi_{sa}$ is included in the simulation, so the qubits start and end in the storage states). A  finite pulse strength, corresponding to a duty cycle $N_p t_p/t_c = 0.025$, is included on all pulses.

Sequences 1 and 2 from Fig. \ref{fig:int_storage}(b,c) attain coherent errors below $10^{-5}$. Further suppression can be achieved by using a sequence that consists of sequence 2 concatenated with its inverse (sequence 3, Fig. \ref{fig:sequences_SI}c), which maintains the robustness of sequence 2 but is reflection symmetric to cancel higher-order terms that arise during the active gate from the non-negligible value of $J_z^{sa}$ ($J_z^{sa}/J_z^{ss} \approx -3.2$). This sequence realizes coherent errors well below $10^{-6}$.

Since the $1/R^6$ interaction is inherently short-ranged, gates can be applied in parallel on multiple pairs of atoms within a large array. In 1D, a separation of $3 a_{circ}$ (\emph{i.e.}, two intermediate storage sites) will result in a cross-talk error below $10^{-6}$. Thus, half of the array can be participating in a gate at any point in time.

\section{Other sources of error}
\label{sec:errors}

In this section, we consider several potentially significant sources of error, estimate their impact on the gate fidelities, and discuss mitigation strategies.

\subsection{Pulse Imperfections}

Imperfections and finite rotation strength in the pulses for the DD sequence can have a significant impact on the fidelity of the DD in the storage states. However, these can be mitigated by careful pulse sequence design. In Fig. \ref{fig:errors}b-c, we show the errors resulting from finite pulse duration (expressed in terms of the pulse duty cycle $N_p t_p/t_c$ where $N_p$ is the number of pulses in one cycle and $t_p$ is the pulse duration) and static rotation angle errors (\emph{i.e.}, resulting from inhomogeneous microwave field strength across the array). Sequences 2 and 3 perform considerably better than sequence 1 for both, and can achieve coherent errors of less than $10^{-6}$ per cycle for duty cycles approaching 1 (equivalent to zero free precession period between the pulses) and rotation angle errors of more than several percent.

\subsection{Atomic motion}
\label{sec:motion}

\begin{figure}
    \includegraphics[width=3 in]{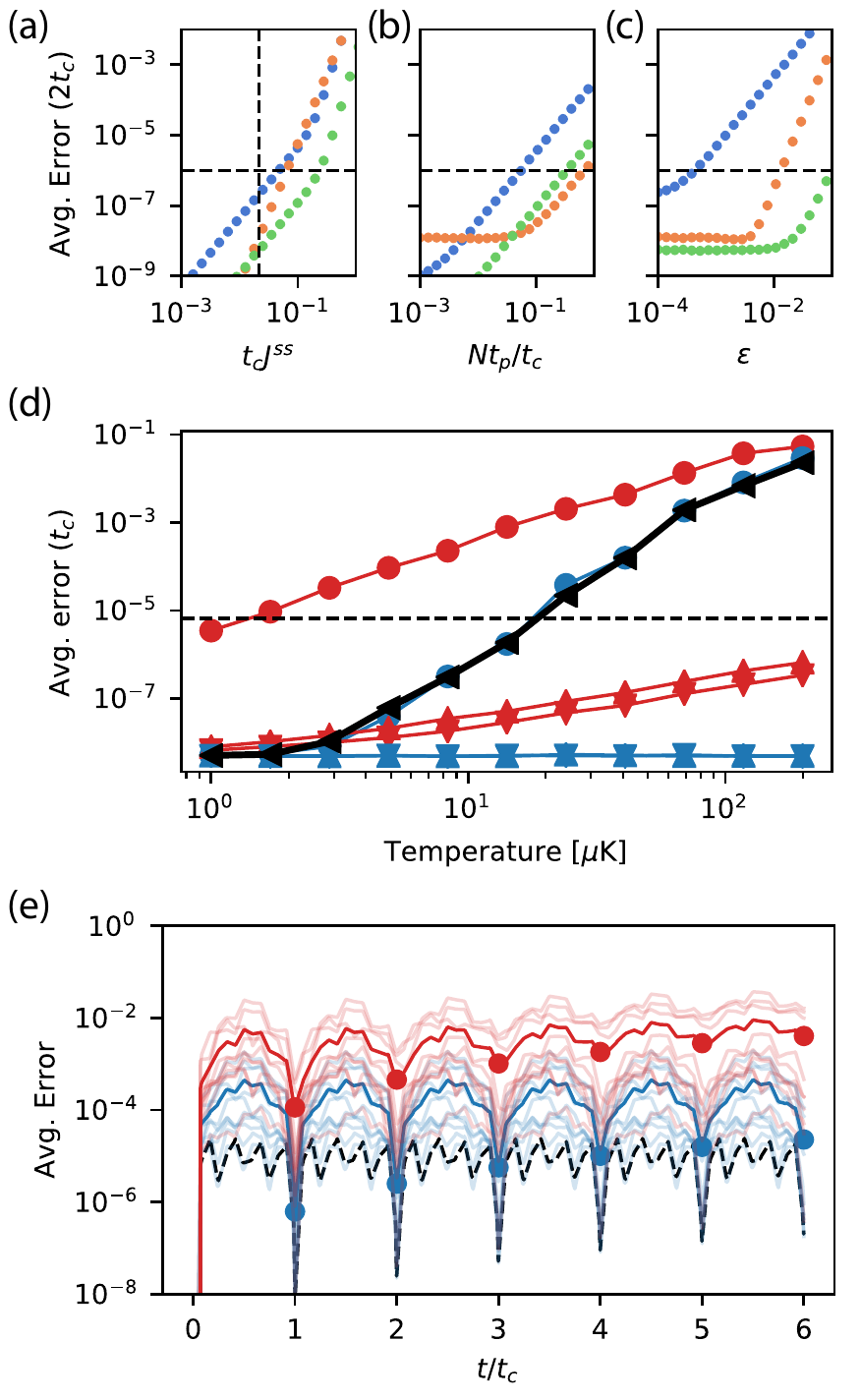}
    \caption{(a-c) Storage state DD errors at $t=2t_c$ for sequences 1-3 as a function of (a) sequence period, (b) pulse duty cycle, and (c) static rotation angle error. Each plot uses the values $(t_c J^{ss}, N_pt_p/t_c, \epsilon) = (0.021,0.025,0)$ for the parameters not being varied. The horizontal line is at $10^{-6}$, a characteristic value for the incoherent error over one two-qubit gate cycle.
    (d) Sensitivity to errors in sequence 2 arising from thermal motion with (blue) and without (red) matching the sequence period to the atomic motion frequency. The red and blue circles ($\bigcirc$) show the effect of a non-magic trap ($\eta = 2 \times 10^{-3}$); up triangles ($\bigtriangleup$) show the effect of the position-dependence of $\Delta^{ss}$; down triangles ($\bigtriangledown$), $J_z^{ss}$ and $J^{ss}$. The black triangles show the sum of all errors at the optimal sequence period. The dashed line shows the lifetime-limited error over one cycle, $2 \times 2 \pi / (\omega \tau_{circ}) \approx 7 \times 10^{-6}$.
    (e) Storage state DD fidelity with no atomic motion (black dashed), $T=10\,\mu$K (blue) and $T=50\,\mu$K, with sequence 2 at the optimal period. Light traces show the evolution for individual randomly sampled trajectories, and the dark traces show the average. In (d,e), $a_{circ} = 16\,\mu$m to match the condition $t_\pi = 4 \times 2 \pi/\omega$ with $\omega = 2\pi\times 100$ kHz.}
    \label{fig:errors}
\end{figure}

A major technical imperfection in all atomic and ion qubit platforms is unwanted atomic motion \cite{Saffman2005,sorensen1999}. It is often the leading source of error in standard Rydberg blockade gates, where it enters as a Doppler shift and variation in the interaction strength \cite{Isenhower2010,Wilk2010,Jau2015,Levine2019,Graham2019} and spin-motion entanglement from photon recoil \cite{robicheaux2021}.

For circular Rydberg qubits, the Doppler shift and photon recoil are negligible for the microwave frequency transitions between circular states. Atomic motion still enters in other ways, chiefly as a variation in the interaction parameters and drive strength, and as a time-dependent energy shift in non-magic traps. However, the fact that the gate operations occur at speeds comparable to the atomic motion allows these effects to be suppressed using dynamical decoupling, effectively exploiting the long correlation time of this noise: a trapped atom is a high-$Q$ mechanical oscillator. Slow heating can arise from photon scattering or trap intensity and position noise, but we note that motional coherence times up to 12 seconds have been observed in optical lattices \cite{ferrari2006}. In this section, we give a conceptual overview of the approach to DD in the presence of thermal motion, and refer the reader to Appendix \ref{app:thermal} for a fully quantum mechanical treatment of spin-motion coupling using average Hamiltonian theory. 

Let us first consider the two-qubit gates. During a two-qubit gate, atomic motion creates an uncertainty in $J^{aa}_z$, which in turn leads to an uncertainty in the accumulated phase. The resulting infidelity is approximately $10^{-3}$ for atoms at $T_a = 10\,\mu$K in a trap with motional frequency $\omega=2\pi \times 100$ kHz. However, $\bar{S}_z$ is constant over the duration of the gate, so the final accumulated phase is given by the average value of $J^{aa}_z$ over the gate time. If the gate time is chosen to satisfy $t_\pi = 2 \pi n / \omega$, the average value will be independent of the motional amplitude $x_0$ along the inter-atomic axis to first order in $x_0/a_{circ}$, and also independent of the phase of the motion with respect to the gate. If $\omega=2\pi \times 100$ kHz, this requires slowing down the gate to $t_\pi = 10\,\mu$s, which will increase the incoherent error during the gate by a factor of 3. Only the motion along the inter-atomic axis is relevant: the motion in the orthogonal directions enters to second order.

In a non-magic trap, atomic motion also gives rise to dephasing. In the active states, a random phase with average magnitude $\sqrt{\langle \phi_D^2 \rangle} = \eta_a k_B T_a t_\pi/(2 \hbar)$ accumulates during the gate \cite{Kuhr2005}, leading to a bit flip probability $P_\phi = \phi_D^2/6$ (here $\eta_a$ is the fractional difference in trap depth for $\ket{0_a}$ and $\ket{1_a}$). For $t_\pi = 10\,\mu$s and $T_a=10\,\mu$K, $P_\phi <10^{-6}$ requires $\eta < 4 \times 10^{-4}$, which can be achieved as described in Section \ref{sec:trapping}. Alternatively, this phase can be cancelled using a composite sequence where the atoms are brought to the active states for a time $t_\pi/2$ (such that a nonlinear phase of $\pi/2$ is accumulated), then returned to the storage states where an $X$ (bit flip) operation is applied before going back to the active states for $t_\pi/2$ again. In this sequence, if $t_\pi/2$ is an integral multiple of the motional period (\emph{i.e.}, $t_\pi = 2 \pi (2n) / \omega$), then the same phase $\phi_D$ is accumulated by both qubit states of each atom, and it becomes a global phase that factors out. We note that motion in all three directions contributes to $\phi_D$, so achieving perfect cancellation requires the three trap frequencies to be matched (or have integer ratios).

Atomic motion also affects the atoms in storage states, resulting in errors during idle operations. However, as we derive in Appendix \ref{app:thermal}, these effects can be mitigated by careful design of the DD sequence, in analogy to the design of filter functions for quantum sensing of time-dependent fields \cite{biercukDynamicalDecouplingSequence2011,degenQuantumSensing2017}. Sequence 2 was designed to decouple from all types of errors (non-magic trapping and variation in the interaction parameters $J,J_z$ and $\Delta$) at commensurate frequencies, such that they can be simultaneously suppressed when $t_c = 2 \times 2\pi/\omega$. This is compatible with the condition above to decouple the active states from atomic motion if $t_\pi/2 = 2 \times 2\pi/\omega$, or equivalently, $t_\pi = 4 \times 2\pi/\omega \approx 40\,\mu$s. In this case, the lifetime-limited error probability per two-qubit gate is approximately $10^{-5}$.

To demonstrate this suppression, we simulate the dynamics of a four atom chain in the storage states with random thermal motion (Fig. \ref{fig:errors}d,e). The motion is treated classically, as a time-dependent variation in the Hamiltonian parameters at a single frequency $\omega$. In Fig. \ref{fig:errors}d, we show that matching the sequence period to the atomic motion results in a dramatic suppression of the errors (defined as in Fig. \ref{fig:int_storage}d). The dominant error arises from non-magic trapping potentials (here, a value of $\eta_s=2 \times 10^{-3}$ is chosen), and a substantial suppression is achieved by matching the sequence period to the motion. For these parameters, coherent errors are below the incoherent error rate (now $\approx 10^{-5}$ per cycle) for temperatures $T_a \leq 20\,\mu$K. In Fig. \ref{fig:errors}e, we show the error over several cycles for $T_a = 10\,\mu$K and $T_a=50\,\mu$K.

Note that meeting the conditions above requires aligning $t_c$, $t_\pi$ and $\omega$. While $t_c$ can be varied arbitrarily with the sequence timing, $\omega$ cannot be varied over a wide range, so $t_\pi$ needs to be adjusted to match. While this can be done using the F\"orster resonance in Fig. \ref{fig:int_active}a, a better approach is to vary the distance between atoms, which maintains the large ratio of $J_z^{aa}/J_z^{ss}$ that allows the DD sequence to perform well.

Lastly, we consider operations that are not part of the dynamical decoupling sequence, such as single-qubit rotations (at the refocusing times) and the storage-active transition $\Pi_{sa}$. These operations are driven by the focused LG beams, and the resulting Rabi frequency is quite sensitive to misalignment and atomic motion. In particular, for the parameters in Fig. \ref{fig:driving}b, $\Omega = \Omega_0 e^{-r^2/2\sigma^2}$ with $\sigma=(107,200,287)$ nm for $\lambda=(532,1064,1550)$ nm for the $\delta n=2$ transition. For an atom at $T_a = 10\,\mu$K, this results in an average rotation angle error $\epsilon=(0.09,0.03,0.015)$. However, assuming the pulses can be applied quickly with respect to the atomic motion (or separated by an integer number of periods), composite pulses such as BB1 can be used, which can suppress static errors to the level of $\epsilon^6$  \cite{cummins2003}. This is sufficient, in principle, to realize rotation errors below $10^{-5}$ for the above-mentioned conditions. Other beam geometries may also help mitigate this effect.

\subsection{Other sources of error}

Another consideration is population leakage to non-circular states. In particular, if the microwave field polarization is not pure or uniform across the array, transitions to elliptical states can occur during the DD sequence. We estimate that a polarization purity of $E_z/E_+ \approx 2.5 \times 10^{-3}$ and $E_-/E_+ \approx 4.2 \times 10^{-2}$ is sufficient to realize population leakage of less than $10^{-6}$ per $\pi/2$ pulse on the storage transitions (here, $E_z$ denotes the $z$-polarized microwave electric field strength, and $E_+/E_-$ denote the $\sigma^{+}/\sigma^-$ strengths). This estimate considers only the matrix elements---the finite detuning of the final states will give additional suppression depending on the overall strength of the rotation. While demanding, this level of polarization purity has been demonstrated for RF fields driving Rydberg atoms using phased antenna arrays \cite{Signoles2014}, and the well-controlled boundary conditions of the waveguide structure are favorable for engineering a similar level of suppression.

Even with perfect polarization of the driving fields, the dipolar interaction mixes the circular states with nearby elliptical states, resulting in weakly allowed transitions separated by several MHz from the intended transitions for the parameters presented here. Leakage to these states can be suppressed using DRAG pulses or multi-frequency drives \cite{Theis2018}. Moving the atoms farther apart also helps dramatically, both by reducing the mixing and decreasing the interaction strength which reduces the required pulse bandwidth. The analysis of these techniques and their interaction with the performance of the DD sequence is beyond the scope of this work.

Lastly, we note that the typical average Hamiltonian theory analysis considers the evolution of two-body spin operators \cite{Waugh1968,Burum1979,Choi2020}. In a many-body system, nonlocal interactions emerge from higher-order terms in the Magnus expansion. It has been noted \cite{Kuwahara2016,Mori2016,Abanin2017,Abanin2017a} that the Magnus expansion does not formally converge for many-body systems with extensive energy, and at long times, the system should approach an infinite temperature state as it absorbs energy from the drive. However, these works have shown that this behavior does not occur before an exponentially large critical time $t^* = \mathcal{O}\left(e^{1/(t_c J)}\right)$, and that before this time, the average Hamiltonian description is nearly exactly correct up to some order $n^*$. We assume that $t^*$ can be made much longer than the duration of the computation by adjusting $t_c$. The exponential dependence of $t^*$ on $t_c$ has recently been experimentally observed in a cold atom system \cite{rubio-abadal2020}.

\section{Discussion}
Several technical comments are in order. First, we consider the choice of atomic species and Rydberg levels. Although the properties of the circular states themselves are completely independent of the atomic species, the atom affects aspects of trapping, cooling and measurement. Alkaline earth atoms can be readily cooled to very low temperatures using narrow intercombination lines \cite{Cooper2018,Norcia2018,Saskin2019}. Furthermore, the optically active ion core allows trapping in standard tweezers \cite{Wilson2019}, albeit with a strong tradeoff between state-sensitivity and heating rate. This may be mitigated by the possibility to also use the ion core for laser cooling without disturbing the circular electron \cite{Teixeira2020}, and narrow-line cooling may be even possible using  electric quadrupole transitions (to $D$ states), with controllable broadening from a repumper \cite{roos1999}. All of the necessary ingredients can also be found in alkali atoms: cooling to very low temperatures has been demonstrated in tweezers \cite{Kaufman2012,Thompson2013} and magic ponderomotive lattices can provide highly state-insensitive Rydberg trapping.

The choice of $n$ is constrained to $n > 50$ by requiring reasonable dimensions for the microwave structure. As $n$ increases, the atoms must be moved farther apart to stay within the perturbative regime of the van der Waals interaction, which increases the demands on the tweezer and imaging optics for the same number of trap sites. The range $n=50-70$ seems ideal.

Second, we observe that the non-resonant nature of the addressing light creates considerable flexibility for advanced photonics integration for scalable addressing, which is a major challenge with current atom and ion experiments \cite{bruzewicz2019}. Working at longer wavelengths like 1064 or even 1550 nm enables a wider range of materials for integrated photonics, and CMOS-compatible grating outcouplers for LG beams have already been demonstrated \cite{Liu2016,Nadovich2016}. It is an additional advantage to be able to work within the wavelength bands of low-noise fiber lasers and amplifiers, and mode-locked lasers may provide a particularly simple path to driving the necessary 60-110 GHz transitions between computational states \cite{Hayes2010}.

Third, the proposed QND detection scheme also provides an interface between circular and ground state qubits that can be used to realize longer-term storage in hyperfine ground states, as well as photonic interconnects between multiple modules \cite{awschalom2021} for large-scale quantum computing systems \cite{Monroe2014}.

Lastly, other qubit encoding and gate schemes are possible. For example, using states with $\Delta n = 1$ to realize two-qubit gates may allow gate times below 1 $\mu$s with long-range ($1/R^3$) interactions, as well as multi-qubit gates using the dipole blockade. Alternatively, it may be possible to exploit the vast multiplicity of elliptical states to realize analogs of bosonic codes \cite{albert2020}. Deliberate introduction of microwave resonances in the structure could enable very long-range interactions between circular atoms in certain states, or controllable dissipation. Finally, given the long atomic lifetime, physically moving qubits within the array may be a viable path towards long-range connectivity.

\section{Conclusion}

We have proposed an architecture for a quantum computer based on individually trapped and manipulated circular Rydberg atoms. Leveraging the seconds-scale lifetimes available in a cryogenic, engineered microwave environment, we anticipate that two-qubit gate errors around $10^{-5}$ are achievable in system sizes of hundreds of atoms without ground state cooling. We have additionally proposed a technique for high fidelity and rapid QND measurements of the circular Rydberg atom state. This is used to overcome low circular state initialization fidelities and for measuring the final circuit output. It can also be applied selectively to measure error syndromes for fault-tolerant quantum computing, or to implement interactive verification protocols \cite{mahadev2018}. Our gate implementation is robust to small variations in the Hamiltonian parameters arising from atomic motion; this approach may also be useful in the context of gates with polar molecules \cite{ni2018}.

We note that the techniques discussed here may also be very valuable for quantum simulation, even without local addressing. The tunability of the circular Rydberg Hamiltonian between the storage states realizes XXZ model over a wide range of parameters \cite{Nguyen2018}, and the ability to apply strong global drives creates many possibilities for Floquet engineering of more exotic phases \cite{harper2020}. Additionally, the ability to perform QND measurements on subsets of the array opens the door to studying the interplay between measurement and coherent evolution, such as measurement-induced phase transitions \cite{skinner2019,ippoliti2021}.

\begin{acknowledgements}
We gratefully acknowledge helpful conversations with Alex Burgers, Michael Gullans, Andrew Houck, Nathalie de Leon, Shuo Ma, Jared Rovny, Sam Saskin, Jack Wilson and Hengyun Zhou. We also acknowledge Hannes Bernien, Lawrence Cheuk, Shimon Kolkowitz and Hannes Pichler for comments on various stages of the manuscript. This work was supported by ARO PECASE (W911NF-18-10215), ONR (N00014-20-1-2426), DARPA ONISQ (W911NF-20-10021) and the Sloan Foundation.
\end{acknowledgements}

\appendix

\section{Details about the waveguide structure}
\label{app:waveguide}

\begin{figure*}
    \centering
    \includegraphics[width=6in]{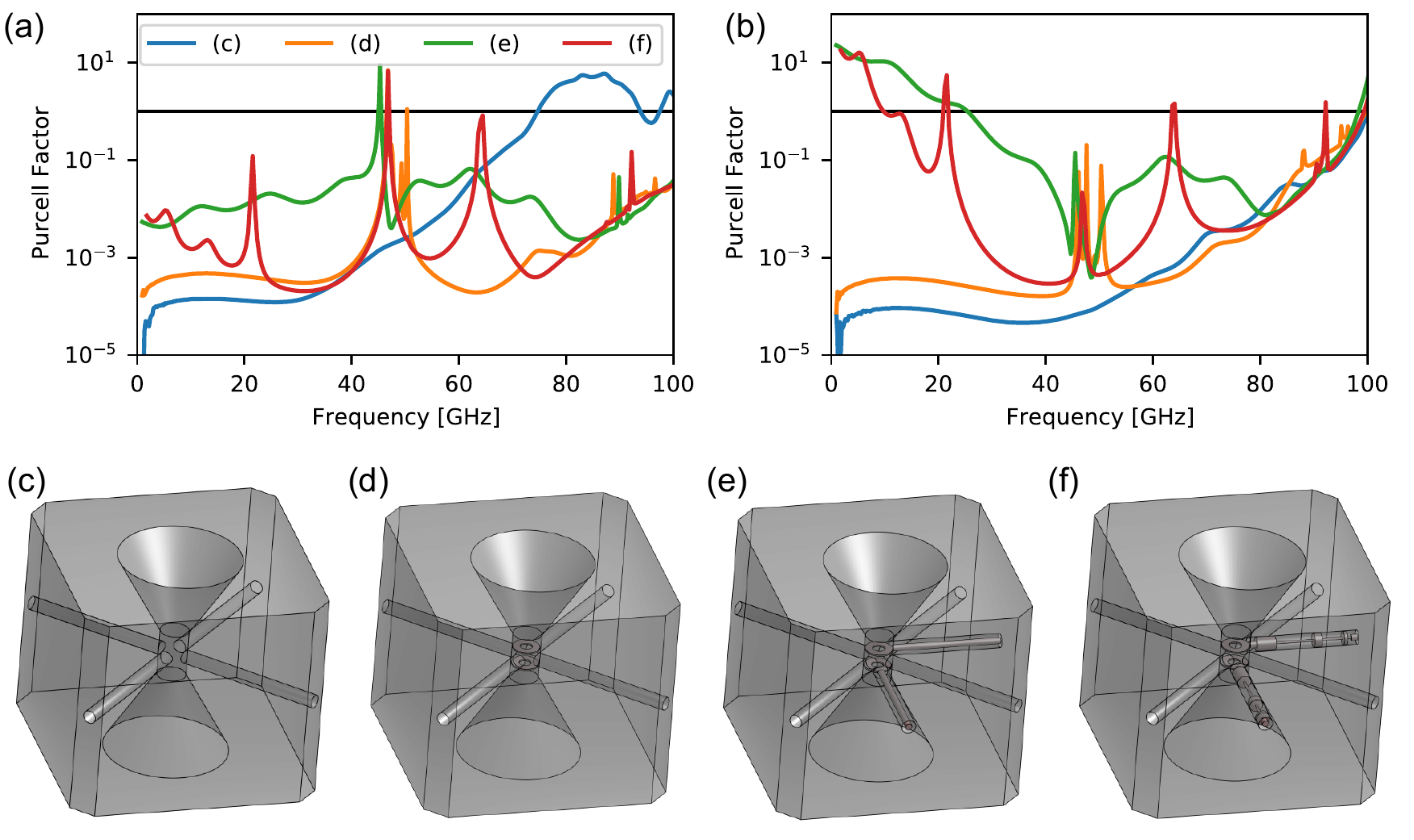}
    \caption{(a) $\sigma^\pm$-polarized Purcell factor and (b) $\hat{z}$-polarized Purcell factor for the microwave structure built up one component at a time. Legend labels refer to illustrations in panels (c-f).
    (c) Waveguide only.
    (d) Waveguide with annular electrodes included.
    (e) Waveguides, annular electrodes and coaxial feedthroughs to apply a bias voltage, terminated with 50 $\Omega$ (not shown).
    (f) Same as panel (e), but with stepped impedance filters added to the feedthroughs. }
    \label{fig:cavity_SI}
\end{figure*}

The center of the waveguide is a cylindrical bore with diameter $D=2.2$ mm. Cones with an opening angle of $60$ degrees (corresponding to an optical NA=$0.5$) are drilled from either side. The apex of each cone is centered at the origin. The electrode rings have a inner and outer diameter of $0.88$ mm and $2.08$ mm and a thickness of $0.1$ mm (the performance is essentially the same with a 0.3 mm thickness), and are separated by $1.43$ mm, chosen to maximize the uniformity of the electric field by zeroing the second derivative with respect to the radial coordinate, leaving only a fourth-order term (the odd terms vanish through the azimuthal symmetry).

The filter is a stepped-impedance filter designed according to an insertion loss method \cite{pozarMicrowaveEngineering2011} for a cutoff frequency of 2.5 GHz. It is housed in a cylinder bore with diameter $1.15$ mm, and the parameters of the inner conductor are given in Table \ref{tab:filter}. We note that a critical property of the filter is that it is highly \emph{reflective}: a dissipative filter would provide a decay channel for the circular states.

\begin{table}
\centering
\begin{tabular}{c|S[table-format=1.3]S[table-format=3.0]S[table-format=1.2]}
    \multicolumn{1}{c}{Segment} & \multicolumn{1}{c}{ID [mm]} & \multicolumn{1}{c}{$Z$ [$\Omega$]} & \multicolumn{1}{c}{$L$ [mm]} \\
    \hline
    1 & 0.48 & 50 & 0.48 \\
    2 & 1.01 & 5 & 1.48 \\
    3 & 0.149 & 120 & 2.99 \\
    4 & 1.01 & 5 & 0.55 \\
    5 & 0.149 & 120 & 1.71 \\
    6 & 1.01 & 5 & 0.26 \\
    7 & 0.149 & 120 & 0.36 \\
    8 & 0.48 & 50 & 0.48 \\
    \hline
    Total & & & 8.29
\end{tabular}
\caption{Parameters of the stepped impedance filter (ID: inner conductor diameter, $Z$: impedance, $L$: length). Segment 1 is the innermost segment.\label{tab:filter}}
\end{table}

The LDOS is simulated using Ansys HFSS with a driven modal solution. A small, perfectly conducting dipole is placed inside, and the radiated power is determined from the real part of the admittance $Y(\omega)$ \cite{reed2010}. The Purcell factor is obtained by normalizing $\Re[Y(\omega)]$ by its value in free space. A radiation boundary condition is used outside the structure, and the filters leading to the electrodes are terminated with a 50 $\Omega$ lumped RLC boundary, representing a resistive terminator thermalized to the bath temperature $T_b$. The conductivity of all components of the structure is taken to be $\sigma=5\times 10^9$ S/m, 100 times that of room temperature copper, appropriate for copper or gold at cryogenic temperatures \cite{jensen1980}.

A simulation of the assembly pieces helps elucidate the contribution of the individual components (Fig. \ref{fig:cavity_SI}). The waveguide structure alone (Fig. \ref{fig:cavity_SI}c) has exponentially decreasing Purcell factor below its lowest cutoff frequency of $f_c = 80$ GHz. It reaches a minimum level of $P_{min} \approx 10^{-4}$ because of the finite electrical conductivity of the sidewalls. This value can be estimated from the Fresnel reflection coefficient $R=\left[(n-1)/(n+1)\right]^2$ for a good metal with complex refractive index $n \approx (1+i)\sqrt{\sigma/2 \omega \epsilon_0}$ as $1-R \approx 4\sqrt{2 \epsilon_0 \omega/\sigma} = 1.3 \times 10^{-4}$ at 50 GHz (Ref. \cite{kleppner1981} contains a similar estimate: $P_{min} = \delta/D$, where $\delta$ is the skin depth). The addition of the annular electrodes (Fig. \ref{fig:cavity_SI}d) results in an additional suppression at high frequencies, presumably because they form effective mirrors for the fundamental $TE_{11}$ mode. It also introduces a resonance at $f \approx 45$ GHz, which may be understood to arise from a series-LC equivalent circuit for a ring inside a waveguide \cite{marcuvitz1951}. Connecting these electrodes to the environment using a 50 $\Omega$ transmission line (Fig. \ref{fig:cavity_SI}e) creates a significant decay pathway, resulting in $P>10^{-2}$ at most frequencies. The effect is larger for $z$-polarization than for in-plane dipoles, presumably because the coupling to the annular electrodes is largely suppressed by symmetry in the latter case. The final addition of the reflective filter (Fig. \ref{fig:cavity_SI}f) suppresses the transmission at most frequencies, but introduces additional resonances near 20 and 60 GHz. We note that these resonances are not completely understood by us, and can be pushed around somewhat by changing the relative dimensions of the components. More exploration may allow a refined design with larger interior space (\emph{i.e.}, more separation between the atoms and the metal walls), or greater optical access.

Finally, we note that the LDOS is essentially constant over a significant volume in the center of the structure. In particular, we observe no significant change in the LDOS for displacements of more than 200 $\mu$m in any direction from the geometric center.

\section{Circular state lifetime}
\label{app:lifetime}
The lifetime of the circular states is affected by a number of processes in addition to spontaneous emission and blackbody radiation. Ref. \cite{Nguyen2018} presents a thorough discussion, which is largely applicable to the present work, including the negligible role of auto- and photo-ionization processes for the circular states and the estimate of background gas collisions in cryogenic vacuum conditions (a transition rate of 1/400 s$^{-1}$ is estimated at $10^{-14}$ torr).

The total radiative lifetime is estimated in Fig. \ref{fig:cavity}, and exceeds 20 seconds at $T_b=4$ K for all of the computational states. However, in ponderomotive optical trap, there is a state transition mechanism resulting from Thomson scattering \cite{Nguyen2018}. A naive estimate of this rate is discussed in section \ref{sec:manip}. Through a detailed calculation \cite{crowley2014}, we have estimated that approximately 2/3 of such scattering events result in a state change, with the vast majority of these causing a transition to one of the neighboring circular states. Therefore, we estimate a state-changing scattering rate of $3\times 10^{-7}$ times the trap depth, or $0.3\,\textrm{s}^{-1}$ for a 1 MHz deep lattice. This is the dominant limitation to the circular lifetime, resulting in $\tau_{circ} \approx 3$ s. We note that this loss rate could be reduced by 10-100 times using near-resonant traps based on the ion core polarizability in alkaline earth atoms, at the expense of a larger motional heating rate from photon scattering in the core.

Lastly, Ref. \cite{Nguyen2018} considers mixing with shorter-lived elliptical states arising from the van der Waals interaction. In our structure with a complete LDOS suppression for all polarizations, the lifetime of the closest few elliptical states is nearly as long as the circular states (the radiative transition rate for states with $|m| = n-2$ is only 3 times greater, for example), and the admixtures are small (less than a few percent) at the larger separations used here. Therefore, this decay process is negligible for our parameters.

\section{Initializing the circular array}
\label{app:loading}

The following procedure can be used to initialize the circular atom array using Rb. Atoms are initially loaded into optical tweezer arrays and excited to $\ket{a}$ and then to $\ket{54D}$ and $\ket{53F}$ using a series of microwave transitions. From there, they are circularized into $\ket{53C}$ using RF rapid adiabatic passage \cite{Nussenzveig1993,Signoles2017}. Then, a series of narrow-band microwave pulses transfers the atoms from $\ket{53C}$ to $\ket{64C}$ ($\ket{1_a}$), where the ancilla array is used to probe which sites were successfully excited. Importantly, the final microwave transfer step leaves behind imperfectly circularized atoms in long-lived elliptical states in the $n=53$ manifold \cite{Signoles2017}. Since the interaction with the ancilla does not distinguish circular atoms from nearly-circular atoms of the same $n$, these would be erroneously recorded as successful initialization if the circularization was performed directly in the $n=64$ manifold. Sites that are not confirmed to be in $n=64$ are emptied by turning off the traps, and the remaining atoms are rearranged into a defect-free array with the desired pattern \cite{Endres2016,Barredo2016,Kim2016}.

The approach outlined above starts with excitation to $\ket{a}$, and therefore has the benefit of requiring only a single Rydberg laser for both excitation and measurement. However, the transfer from $\ket{53C}$ to $\ket{64C}$ requires a large number of microwave frequencies. This could be circumvented by exciting directly to $63S$, from which one would circularize to $\ket{1_s}$, at the expense of needing a second Rydberg excitation laser.

If the compute atoms are to be held in a ponderomotive lattice, several stages of traps are required to initialize the array. First, compute and ancilla arrays are prepared in standard, red-detuned optical tweezers, and rearranged using existing techniques demonstrated in 1D \cite{Endres2016}, 2D \cite{Kim2016,Barredo2016} and 3D \cite{Barredo2020}. Then, a shallow-angle blue-detuned lattice is applied along the vertical direction, confining the two arrays in planes separated by $d_z$ (the combination of a vertical lattice and tweezers has been recently demonstrated \cite{young2020}). Next, the compute array is transferred into a set of hollow, blue-detuned tweezers formed by $LG_{0,1}$ beams that can confine both ground state and Rydberg atoms (the vertical confinement is provided by the lattice) and excited into circular states as described above. The atoms are measured using the ancilla array, and then the compute array rearranged using the $LG_{0,1}$ tweezers. Finally, the compute and ancilla atoms are transferred to the in-plane, state-insensitive lattice before the computation starts. The in-plane lattice will have the opposite polarizabilty for the ground state ancillae compared to the circular states, but trapping the ancilla array with a lateral offset of $\lambda/2 \approx 300$ nm will have negligible impact.

The sequence is somewhat simpler using alkaline earth atoms and near-resonant optical tweezers for the circular states. The compute and ancilla arrays are initialized in red-detuned tweezers and rearranged. Then, the compute array is excited to circular states following a similar sequence, though the details depend on the atomic species \cite{Teixeira2020}. For an appropriately chosen tweezer wavelength and beam waist, the low-$\ell$ and circular states can be trapped in the same tweezer that confines the ground state atoms \cite{Wilson2019}. The circular excitation is verified with the ancilla array, and the compute array is rearranged a second time. Finally, the defect-free circular atom array is transferred into a superimposed near-resonant tweezer array that provides state-insensitive trapping, keeping the ancilla array in ground-state tweezers. If out of focus light from the ancilla traps adversely affects the compute array, the ancillae may be moved away during the computation or switched into a configuration where they are in the same plane as the compute array but displaced laterally by $d_x = a_{circ}/3$. The resulting interactions are essentially the same as those in Fig. \ref{fig:detect}.

\section{Measurement fidelity}
\label{app:meas}

Here we consider the fidelity of the measurement process in detail. For concreteness, we assume a lifetime of the ancilla $\ket{S}$ Rydberg state of $\tau_a = 200\,\mu$s (appropriate for $\ket{55S}$ at cryogenic temperatures), and assume that the interaction strength between an ancilla and a target atom in the target state is $\Delta_{t} = V_{a,64} = 2\pi \times 20$ MHz (Fig. \ref{fig:detect}).

\begin{figure}
    \centering
    \includegraphics[width=3.5 in]{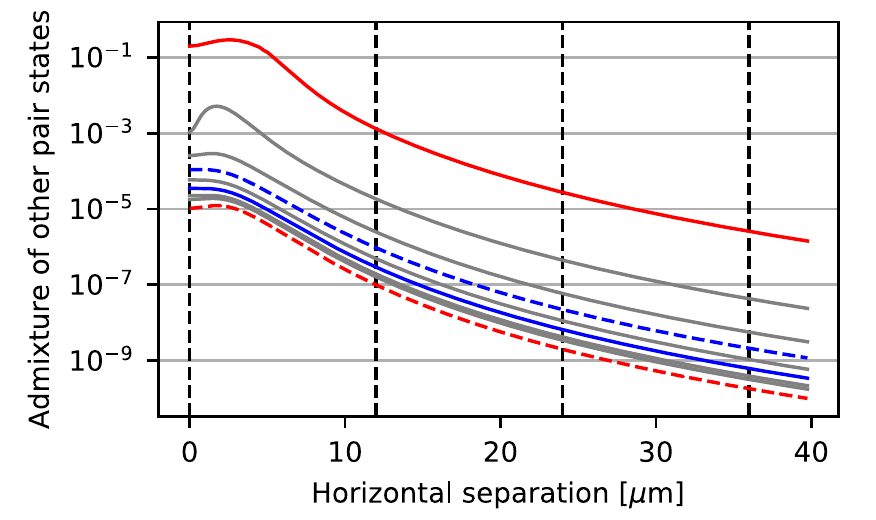}
    \caption{Admixture of other pair states $P_\epsilon$ into the pair $\ket{a,nC}$ for $n=56-66$. The computational states are color-coded following Fig. \ref{fig:fig1}a, and the others are shown in gray. The horizontal lines show lattice sites with $a_{circ}=12\,\mu$m.}
    \label{fig:meas_SI}
\end{figure}

One approach to reading out the circular state is to execute a Rydberg blockade gate on the ancilla. As with a conventional Rydberg blockade gate \cite{Lukin2001}, a $2\pi$ pulse on the ancilla atom with Rabi frequency $\Omega_a$ creates a $\pi$ phase shift if it is not blockaded by the target atom. In this approach, errors can arise from the finite blockade strength as well as the finite lifetime of the circular Rydberg state, and are minimized at an optimum value of $\tilde{\Omega}_a = (\pi \Delta_t^2/\tau_a)^{1/3} \approx 2 \pi \times 1$ MHz. At this value, the error probability is $P_g = \left[\pi/(\Delta_t \tau_a)\right]^{2/3}/2 \approx 1.3 \times 10^{-3}$ (this is the same error scaling as a two-atom Rydberg blockade gate \cite{Saffman2016}, but with a smaller prefactor since the circular atom has negligible decay over the gate).

Another source of error arises from the possibility for the measurement process to alter the state of the target atom, leaving it in the incorrect state. The van der Waals interaction mixes the single-atom eigenstates of the ancilla and target atom ($\ket{a}$ and $\ket{nC}$) with other states. This results in pair eigenstates of the form: $\ket{\psi_{an}} = \sqrt{1-\sum_i |\epsilon_{n,i}|^2} \ket{a,nC} + \sum_i \epsilon_{n,i} \ket{a'_i,n'_i}$, where $\ket{a'_i,n'_i}$ are other states that are mixed in by the dipolar interaction (the states $\ket{a'_i}$ are all $P$ states, while the states $\ket{n'_i}$ are circular or nearly-circular). A spontaneous decay of the ancilla atom has a probability $P_\epsilon(n) = \sum_i |\epsilon_{n,i}|^2$ of projecting the target atom into a different state (assuming the lifetimes of $\ket{a}$ and $\ket{a'_i}$ are similar). In Fig. \ref{fig:meas_SI}, we show this quantity for various circular states.

To estimate the error resulting from this effect, we consider the blockaded and non-blockaded cases separately. In the blockaded case ($n=64$), the $P_\epsilon \approx 0.2$ is rather large, but the spontaneous emission probability $P_{sc}$ is very small because the ancilla is only excited with low probability, $P_{sc} = \left[\pi / (\Delta_t \tau_a)\right]^{4/3} \approx 6.3 \times 10^{-6}$. This gives a total pair projection error probability $P_p = P_\epsilon P_{sc} = 1.2 \times 10^{-7}$. In the other states ($n=56,59,61$), $P_\epsilon \lesssim 10^{-4}$, but the probability to decay is much higher, $P_{sc} = P_g/2$, resulting in $P_p \lesssim 10^{-7}$. In both cases, these errors are very small compared to $P_g$. One can also consider cross-talk: the probability for an ancilla to project its target atom's \emph{neighbor} is, in the worst case, $P^{NN}_p = (P_g/2) P_\epsilon^{NN} \approx 10^{-6}$, where $P_\epsilon^{NN} \approx 10^{-3}$ is evaluated at the nearest neighbor distance for an atom in $\ket{64C}$ (Fig. \ref{fig:meas_SI}). This is also small.

This measurement technique is also suited to QND readout of a qubit array, where it is desired to maintain an arbitrary superposition of $\ket{0_s},\ket{1_s}$ in a neighboring atom. For these states at the nearest-neighbor distance, $P_\epsilon < 10^{-6}$. However, there will be a small phase rotation resulting from the differential interaction energy of the storage states with the excited neighboring ancilla. The ancilla Rydberg state is populated for an average time $2 \pi / (8 \tilde{\Omega}_a) = 0.13 \,\mu$s, and the differential shift on a neighboring atom in $\ket{0_s},\ket{1_s}$ is approximately 1 kHz (Fig. \ref{fig:detect}d); therefore, the resulting phase shift is $\phi = 8 \times 10^{-4}$, corresponding to an error probability $P_{\phi} = 1.1 \times 10^{-7}$, much less than that arising from spontaneous emission over the duration of the measurement, $P_d$.

In summary, it is possible to realize measurement errors at the level $10^{-3}$ using a blockade gate on an ancilla atom. The measurement perturbs the target and neighboring atoms at a level below $10^{-6}$, which allows repeating the measurement with several ancillae or with the same ancilla, sequentially, to achieve even higher fidelity by averaging over several repetitions of the gate \cite{hume2007}.

\section{Interaction calculations}
\label{app:interactions}

\begin{figure*}[t]
    \centering
    \includegraphics[width=7in]{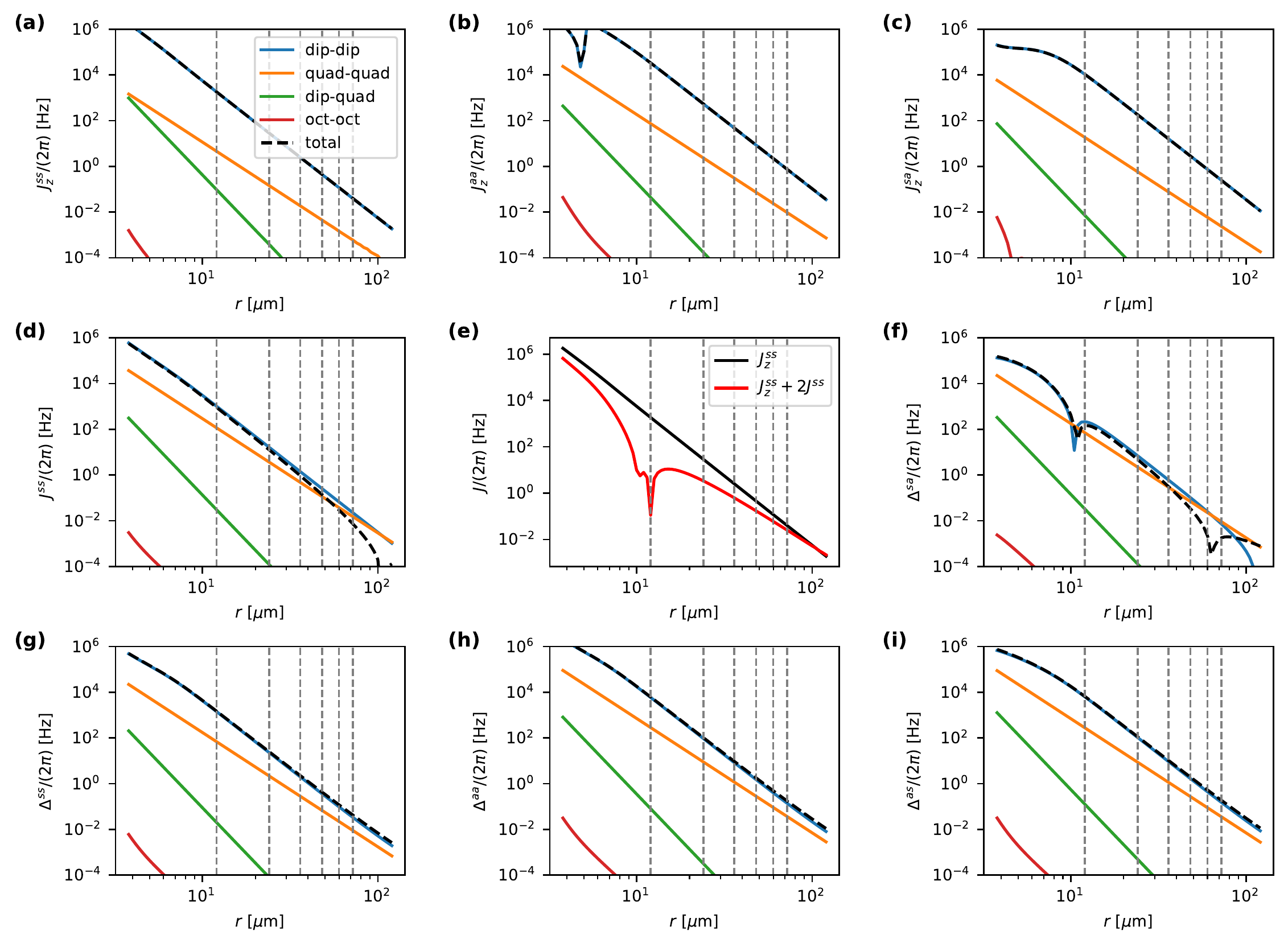}
    \caption{(a,d,g) Magnitude of the interaction coefficients vs. distance for the storage states [Eq. \eqref{eq:Hss}]. The vertical lines show the site separation in a 1D lattice with $a_c = 12\,\mu$m. The colors show the separate contributions from different terms in the multipole expansion. (e) $J_z^{ss} + 2J^{ss}$ as a function of distance. The $B-$ and $E-$fields are tuned to null this quantity precisely for the nearest neighbor site, but the quadrupole contribution to $J^{ss}$ results in imperfect cancellation at more distant sites. (b,h) Interaction coefficients for the active states [Eq. \eqref{eq:Haa}]. (c,f,i) Interaction coefficients between the storage and active states [Eq. \eqref{eq:Hsa}]}
    \label{fig:multipole}
\end{figure*}

The effective interaction coefficients in the Hamiltonians Eqs. \eqref{eq:Hss}-\eqref{eq:Haa} are numerically computed. To compute the interaction between atoms in the circular states $n$ and $n'$, we construct a large basis of pairs of nearby Rydberg states (typically of order $10^3$ pairs are included), and compute a Hamiltonian with one-atom terms (Rydberg state energy, and the effects of $E$ and $B$ fields) and two-atom interaction matrix elements up to the desired multipole order \cite{weber2017} (radial matrix elements are computed using analytic hydrogenic wavefunctions). This matrix is diagonalized, and the pair eigenstates with the highest overlap with the pure $n,n'$ pair state are extracted. If $n=n'$, the van der Waals coefficient $V_{nn}$ is extracted as the shift of the energy of this state with respect to the one-atom Hamiltonian (or, equivalently, with respect to its energy at infinite separation in the pair Hamiltonian). If $n\neq n'$, then we instead find the eigenstates $\ket{\psi_\pm}$ with the highest overlap with the pair states $(\ket{n,n'} \pm \ket{n',n})/\sqrt{2}$. The van der Waals coefficient $V_{nn'}$ extracted from the average energy shift of these states $(E_+ + E_-)/2$, while the exchange interaction coefficient $E_{nn'}=E_+ - E_-$.

If we now associate the state $n$ with spin down, and $n'$ with spin up for a fictitious spin-1/2, we can write the coefficients as a matrix in the product basis and express it in terms of spin operators:

\begin{equation}
\begin{split}
    H_{int} &=
    \left(\begin{tabular}{cccc}
         $V_{nn}$ & 0 & 0 & 0  \\
         0 & $V_{nn'}$ & $E_{nn'}$ & 0 \\ 
         0 & $E_{nn'}$ & $V_{n'n}$ & 0 \\
         0 & 0 & 0 & $V_{n'n'}$
    \end{tabular}\right) \\
    &= J_z S^1_z S^2_z + J(S^1_x S^2_x + S^1_y S^2_y) + \Delta (S^1_z + S^2_z) + E_0 \hat{\mathbb{I}}
\end{split}
\end{equation}

with the operator coefficients (note that $S_z$ has eigenvalues $\pm 1$):

\begin{eqnarray}
J_z &=& (V_{nn} - V_{nn'} - V_{n'n} + V_{n'n'})/4\\
J &=& E_{nn'}/2 \\
\Delta &=& (V_{nn} - V_{n'n'})/4 \\
E_0 &=& (V_{nn} + V_{nn'} + V_{n'n} + V_{n'n'})/4
\end{eqnarray}

We apply this procedure with $(n,n') = (59,61)$ to compute the storage state interactions [\emph{i.e.}, the coefficients $J^{ss}$, $J_z^{ss}$ and $\Delta^{ss}$ in Eq. \eqref{eq:Hss}], and with $(n,n') = (56,64)$ to compute the active state interactions [\emph{i.e.}, $J_z^{aa}$, $\Delta^{aa}$ in Eq. \eqref{eq:Haa}].

To compute the interactions between the storage and active atoms, we consider the  basis implied by the operator $S_z \bar{S}_z$:

\begin{equation}
\begin{split}
H_{int} &=
    \left(\begin{tabular}{cccc}
         $V_{0_s0_a}$ & 0 & 0 & 0  \\
         0 & $V_{0_s1_a}$ & 0 & 0 \\ 
         0 & 0 & $V_{1_s0_a}$ & 0 \\
         0 & 0 & 0 & $V_{1_s1_a}$
    \end{tabular}\right) \\
    &=  J^{sa}_z S^1_z \bar{S}^2_z + \Delta^{sa} S^1_z + \Delta^{as} \bar{S}^2_z + E_0 \hat{\mathbb{I}}
\end{split}
\end{equation}

The operator coefficients are:

\begin{eqnarray}
J^{sa}_z &=& (V_{0_s0_a} - V_{0_s1_a} - V_{1_s0_a} + V_{1_s1_a})/4\\
\Delta^{sa} &=& (V_{0_s0_a} + V_{0_s1_a} - V_{1_s0_a} - V_{1_s1_a})/4 \\
\Delta^{as} &=& (V_{0_s0_a} - V_{0_s1_a} + V_{1_s0_a} - V_{1_s1_a})/4 \\
E_0 &=& (V_{0_s0_a} + V_{0_s1_a} + V_{1_s0_a} + V_{1_s1_a})/4
\end{eqnarray}

\begin{table}[tp]
    \centering
    \begin{tabular}{c|S[table-format=+2.3]}
    \multicolumn{1}{c}{Term} & \multicolumn{1}{c}{Value ($2 \pi \times 10^{3}$ s$^{-1}$)} \\
    \hline
     $J^{ss}$ & -0.918 \\
     $J_z^{ss}$ & +1.84 \\
     $J_z^{sa}$ &  -10.62 \\
     $J_z^{aa}$ & 33.61 \\
     $\Delta^{ss}$ & -1.51 \\
     $\Delta^{sa}$ & 0.144 \\
     $\Delta^{as}$ & -6.60 \\
     $\Delta^{aa}$ & -6.24 \\
    \end{tabular}
    \caption{Computed values of the interaction coefficients at $a_{circ}=12\,\mu$m, with $E_z = 0.313$ V/cm and $B_z = 1.39$ G.}
    \label{tab:terms}
\end{table}

In Fig. \ref{fig:multipole}, we plot the value of the operator coefficients as a function of distance, and their values at $a_{circ} = 12\,\mu$m are given in Table \ref{tab:terms}. We have tuned in the condition $J_z^{ss} = -2J_{ss}$ with $E_z = 0.313$ V/cm and $B_z = 1.39$ G. We include higher multipoles than dipole-dipole, and observe that the quadrupole-quadrupole interaction (with $1/R^5$ dependence) contributes somewhat significantly to $J^{ss}$ and $\Delta^{sa}$. The main consequence of this is that the dipolar condtion $J^{ss} = - 2 J_z^{ss}$ can only be exactly satisfied at single distance, as seen in Fig. \ref{fig:multipole}e. However, the error at the next-nearest neighbor site is only a few Hz, which can be expressed as an extra $J_z$ contribution (note that this is not included in the simulations in Figs. \ref{fig:int_storage} and \ref{fig:int_active}). If the effect of this term is significant, it can be removed with a slow, local dynamical decoupling sequence.

Eqs. \eqref{eq:Hss} - \eqref{eq:Haa} describe the effect of interactions on atoms in arbitrary superpositions of the two storage state levels and arbitrary superpositions of the two active levels. They do not capture superpositions of storage and active levels, however, these types of states should not occur in the protocol. There are interaction terms coupling these states that are not represented in the effective operator Hamiltonian. However, these are all higher-order than quadrupole-quadrupole, and as seen in Fig. \ref{fig:multipole}, the magnitude of these terms is negligible. The largest term not included in Eqs. \eqref{eq:Hss} - \eqref{eq:Haa} is an exchange term between $\ket{1_s}$ and $\ket{1_a}$, which has a magnitude $E_{1_s,1_a} = -0.9$ Hz at 12 $\mu$m separation (it arises primarily from the third-order dipole-dipole interaction with $1/R^9$ dependence).

A similar approach is used to calculate the interactions between the ancilla states and the circular states. The only difference is that non-hydrogenic wavefunctions are used for the low-$\ell$ ancilla states, incorporating the finite value of the quantum defect \cite{weber2017}.

\onecolumngrid
\section{Decoupling from atomic motion}
\label{app:thermal}
To evaluate the impact of atomic motion on the active and storage operation fidelities, we construct an average hamiltonian theory (AHT) model incorporating the motion as a quantum degree of freedom. We begin by considering the storage states, writing the Hamiltonian $H_s = H_{m} + H_{\eta} + H_{int}$, with:

\begin{align}
    H_m &= \sum_i \frac{\hat{p}_i^2}{2m} + \frac{1}{2} m \overline{\omega^2} \hat{x_i}^2 \\
    H_\eta &= \sum_i \frac{1}{2} \eta' m \overline{\omega^2} \hat{x_i}^2 S_z^i \\
    H_{int} &= \sum_{ij} J^{ij}_z f(\hat{x}_{ij}) S_z^i S_z^j + J^{ij} f(\hat{x}_{ij}) \left(S_x^i S_x^j + S_y^i S_y^j \right) + \Delta^{ij} f(\hat{x}_{ij}) S_z^i n^j
\end{align}

$H_m$ describes the motion of the trapped atom with coordinates $\hat{x}_i,\hat{p}_i$ ($x_i$ is defined relative to the trap center). $H_\eta$ describes the state-dependent potential resulting from a non-magic trap. If the potential for the atom in the state $S_z=-1$ is $U_0 = \frac{1}{2}m \omega^2 x^2$, and the potential for the atom when $S_z=1$ is $U_1 = (1+\eta) U_0$ (following Section \ref{sec:trapping}), then $\overline{\omega^2} = (1+\eta') \omega^2$ with $\eta' = \eta/2$. $H_{int}$ describes the interaction of pairs of atoms in the storage states [following Eq. \eqref{eq:Hss}, but with $ss$ superscripts removed for simplicity]. The coefficients $J_z^{ij}$, $J^{ij}$ and $\Delta^{ij}$ have an implicit dependence on the average separation as $1/r_{ij}^6$. In subsequent steps, we will approximate the distance-dependence of the interaction terms to first order, as $f(\hat{x}_{ij}) = 1-6\hat{x}_{ij}/r_{ij}$ [$r_{ij}=(i-j)a_{circ}$ is the average separation, and $\hat{x}_{ij} = \hat{x}_i - \hat{x}_j$].

This system is then acted on by a series of pulses in the DD sequence. For simplicity, we take these pulses to be equidistant occurring at times $t_k = k \tau$, and let $P_k$ denote the unitary transformation (on the spin) realized by the $k^th$ pulse. Following a standard AHT treatment \cite{Haeberlen}, we define a toggling frame Hamiltonian representing the evolution of the spin during the time interval between $t_{k-1}$ and $t_k$:

\begin{equation}
    \tilde{H}_k = (P_{k-1}...P_1)^\dag H_s (P_{k-1}...P_1)
\end{equation}

Additionally, we go into an interaction picture $\hat{H}_k$ with respect to $H_m$: $\hat{H}_k(t) = e^{-i H_m t} \tilde{H}_k e^{i H_m t}$. This results in the elimination of $H_m$, as well as the following substitutions that make $\hat{H}_k(t)$ explicitly time-dependent:

\begin{align}
    \hat{x}_i &\rightarrow x_0 \left( e^{-i \bar{\omega} t} a_i + e^{i \bar{\omega} t} a_i^\dag \right) \\
    \hat{x}^2_i &\rightarrow x_0^2 \left( e^{-2 i \bar{\omega} t} a_i^2 + e^{2 i \bar{\omega} t} (a_i^\dag)^2 + 2 a_i^\dag a_i + 1 \right) \\
    \hat{x}_{ij} &\rightarrow x_0 \left( e^{-i \bar{\omega} t} a_{ij} + e^{i \bar{\omega} t} a_{ij}^\dag \right)
\end{align}

with $a_{ij} = a_i - a_j$. Note that $a_i$ is not time dependent.

If the resulting Hamiltonian $\hat{H}_k(t)$ is periodic in the sense that it returns to itself after $N$ pulses and a time $T=N\tau$ [\emph{i.e.}, $\hat{H}_k(t) = \hat{H}_{k+N}(t+T)$], then the propagator has the property $U(M T) = U(T)^M$ for integer $M$. In that case, the system dynamics at the refocusing times $t=MT$ can be usefully approximated with a time-independent Hamiltonian using the Magnus expansion \cite{Haeberlen}. Specifically, we can approximate $U(T) = e^{i H_{\textrm{eff}} T}$ with $H_{\textrm{eff}} = \bar{H}^{(0)} + \bar{H}^{(1)} +  ...$. The first terms are:

\begin{align}
    \bar{H}^{(0)} &= \frac{1}{T} \int_0^T  dt_1 \hat{H}(t_1) \\
    \bar{H}^{(1)} &= -\frac{i}{2T} \int_0^T dt_2 \int_{0}^{t_2} dt_1 [\hat{H}(t_2),\hat{H}(t_1)]
\end{align}

We will only analyze $\bar{H}^{(0)}$ in the context of spin-motion coupling, although we note that some of the presented DD sequences have a significant cancellation of $\bar{H}^{(1)}$ in the absence of motion. In fact, by virtue of its reflection symmetry, sequence 3 has no contribution from any odd-order term \cite{Burum1979}.

This periodicity in $\hat{H}_k(t)$ imposes two conditions: that the pulses transform the spin operators back to themselves after $N$ pulses, and that the frequency of the atomic motion $\bar{\omega}$ satisfies $\bar{\omega} T = 2 \pi n$ for some integer $n$. Since the frequency of the atomic motion is known, the second condition can always be accomplished by varying the pulse spacing, $\tau$.

Following the notation in Ref. \cite{Choi2020}, we represent the action of the driving pulses on the spin operators in $\hat{H}_k$ using a matrix $F_{\mu,k}$ that represents the transformation of the $S_z$ operator after the $k^{th}$ pulse (a graphical depiction of $F_{\mu,k}$ is given in Fig. \ref{fig:sequences_SI}):

\begin{align}
    S_z^i &\rightarrow \sum_{\mu} F_{\mu,k} S_\mu^i \\
    S_z^i S_z^j &\rightarrow \sum_{\mu} F_{\mu,k}^2 S_\mu^i S_\mu^j \\ 
    (S_x^i S_x^j + S_y^i S_y^j) &\rightarrow \sum_{\mu} (1-F_{\mu,k}^2) S_\mu^i S_\mu^j
\end{align}

We now examine the contributions to $\bar{H}^{(0)}$ from each term in $H_s$. The conditions to cancel their contributions are derived and summarized in Table \ref{tab:conditions}.

\subsection{Disorder terms}

We now consider the zeroth-order AHT for the terms in $H_s$ that depend only on $S_z^i$, which we call $\bar{H}_{dis}^{(0)}$ since they act as a disorder in the local field at each site. First, let us write the toggling frame, interaction picture Hamiltonian $\hat{H}_{k,dis}$:

\begin{equation}
    \hat{H}_{k,dis}(t) = \sum_{i,\mu} \frac{1}{2} \eta' m \overline{\omega^2} \hat{x}^2_i(t)F_{\mu,k} S_\mu^i + \sum_{i,j,\mu} \Delta^{ij} \left[1 - \alpha \hat{x}_{ij}(t) \right]F_{\mu,k} S_\mu^i n^j
\end{equation}

with $\alpha = 6/r_{ij}$. Then:

\begin{align}
\label{eq:Hdis0_all}
    \bar{H}_{dis}^{(0)} = \frac{1}{T} \sum_k \int_{(k-1)\tau}^{k \tau} \hat{H}_{k,dis}(t) = \sum_{i,j} \sum_{k,\mu} H^{i,j}_{k,\mu}
\end{align}

with

\begin{align}
\label{eq:Hdis0}
    \begin{split}
    H^{i,j}_{k,\mu} &= \Delta_{ij} F_{\mu,k} S_\mu^i n^j
    + \frac{1}{2} \eta' m \overline{\omega^2} (2 a_i^\dag a_i + 1) F_{\mu,k} S_\mu^i \\
    &- i \frac{\alpha x_0}{\bar{\omega} T} \Delta^{ij} \left[(1-e^{i \bar{\omega}\tau}) e^{-i \bar{\omega} k \tau} F_{\mu,k} a_{ij} - h.c.\right] S_\mu^i n^j
    + i \frac{\eta' m \overline{\omega^2} x_0^2}{4 \bar{\omega} T} \Delta^{ij} \left[(1-e^{2i \bar{\omega}\tau}) e^{-2i \bar{\omega} k \tau} F_{\mu,k} a_{i}^2 - h.c.\right] S_\mu^i
\end{split}
\end{align}

The first term in Eq. \eqref{eq:Hdis0} vanishes after the sum in Eq. \eqref{eq:Hdis0_all} if $\sum_k F_{\mu,k} = 0$ for each $\mu$ in $\{x,y,z\}$, which is the standard condition to decouple from static disorder (Ref. \cite{Choi2020}, and condition 1 in Table \ref{tab:conditions}). The second term reflects the average energy shift in the non-magic trap, and vanishes under the same condition.  The third term can vanish under two distinct circumstances. The first is $\bar{\omega} \tau = 2 \pi n$ (\emph{i.e.}, applying one pulse per motional period). In this case, the position-dependent spin terms average to zero in between each pulse. The second possibility is:

\begin{equation}
    \sum_{k=1}^N e^{i k \bar{\omega} \tau} F_{\mu,k} = 0, \forall \mu
\end{equation}

In this case, the position-dependent spin terms do not average to zero between every pulse, but instead average to zero over the sequence of $N$ pulses (condition 4 in Table \ref{tab:conditions}). The fourth term in Eq. \eqref{eq:Hdis0} vanishes under the same conditions as the third, but with $\bar{\omega} \rightarrow 2\bar{\omega}$ (condition 5 in Table \ref{tab:conditions}).

\subsection{Interaction terms}

Now we can repeat the same calculation for the terms in $H_s$ with two spin operators, which we call $\bar{H}_{int}^{(0)}$. The toggling frame Hamiltonian is:

\begin{equation}
    \hat{H}_{k,int}(t) = \sum_{i,j} \sum_{\mu} J^{ij}_z \left[1 - \alpha_z \hat{x}_{ij}(t)\right] F_{\mu,k}^2 S_\mu^i S_\mu^i 
    + J^{ij} \left[1 - \alpha_J \hat{x}_{ij}(t)\right] (1-F_{\mu,k}^2) S_\mu^i S_\mu^i
\end{equation}

Proceeding as before, we arrive at:

\begin{equation}
    \begin{split}
    \bar{H}_{int}^{(0)} &= \sum_{i,j} \sum_{\mu,k} \left[J^{ij}_z F_{\mu,k}^2 + J^{ij} (1-F_{\mu,k}^2)\right] S_\mu^i S_\mu^j \\
    &- i\frac{x_0}{\bar{\omega}T}\sum_{i,j}  \sum_{\mu,k} \left[J^{ij} \alpha_J (1-e^{i \bar{\omega}\tau}) e^{-i \omega k \tau} a_{ij} +  (J_z^{ij} \alpha_z - J^{ij} \alpha_J)(1 - e^{i \bar{\omega} \tau}) e^{-i \bar{\omega} k \tau} F_{\mu,k}^2 a_{ij} - h.c. \right]S_\mu^i S_\mu^j
    \end{split}
\end{equation}

The first sum describes the motion-independent interactions. When $\sum_{k} F_{\mu,k}^2 = N/3$ for all $\mu$, then the sum is proportional to $(J_z + 2 J)$ and vanishes when the dipolar condition $J_z = -2 J$ is met (condition 2 in Table \ref{tab:conditions}). The first term in the second sum is independent of $F_{\mu,k}$ and vanishes if $\tau \bar{\omega} = 2\pi/n$, which is guaranteed by the periodicity of $\hat{H}$. The final term can vanish under two separate conditions, as in the case of $\bar{H}_{dis}^{(0)}$: if $\tau \bar{\omega} = 2 \pi n$, or if

\begin{equation}
    \sum_{k=1}^{N} e^{-i \bar{\omega}k \tau} F_{\mu,k}^2 = 0, \forall \mu
\end{equation}

This is condition 6 in Table \ref{tab:conditions}.

\subsection{Pulse errors}

Lastly, we consider the effects of pulse angle errors resulting from the atomic motion. While these are negligible for microwave-driven transitions, they would arise if the DD sequence was executed using focused LG modes, and we include this section for completeness. The $k^{th}$ pulse is generated by a Hamiltonian:

\begin{equation}
\hat{H}_{d,k} = \sum_{i,\mu} \Omega \left(1 +\epsilon_i+ \beta_1 \hat{x}_i(t) + \beta_2 \hat{x}_i^2(t) \right) \beta_{\mu,k} S_\mu^i
\end{equation}

Here, $\vec{\beta}_k = \vec{F}_{k+1} \times \vec{F}_k$ represents the rotation axis in the $k^{th}$ toggling frame. The linear term $\beta_1$ arises from misalignment between the trap center and the addressing beam, while the quadratic term $\beta_2$ results from the finite extent (curvature) of the addressing beam. The term $\epsilon_i$ reflects a static rotation angle error, arising from a gradient in the microwave field strength across the array.

Assuming the pulse duration $t_p = \pi/(4 \Omega)$ is much less than $\tau$ and $\bar{\omega}$, the leading pulse error is described by the following contribution to the lowest order average Hamiltonian \cite{Choi2020}:

\begin{equation}
\bar{H}_{dr}^{(0)} = \frac{1}{T} \sum_{i,\mu,k} \left[ \epsilon_i + \beta_1 e^{-i \bar{\omega} k \tau} a_i + \beta_2 e^{-2 i \bar{\omega} k \tau} a_i^2 + h.c. \right] \beta_{\mu,k} S_\mu^i
\end{equation}

The static error term, $\epsilon$, will vanish if $\sum_k \beta_{\mu,k} = 0$ for all $\mu$, which is condition 3 in Table \ref{tab:conditions} \cite{Choi2020}. The term proportional to $\beta_1$ will vanish if $\sum_k e^{-i k \bar{\omega}\tau} \beta_{\mu,k} = 0$ for all $\mu$. The term proportional to $\beta_2$ vanishes under the same condition, but with $\bar{\omega} \rightarrow 2\bar{\omega}$. These are conditions 7 and 8 in Table \ref{tab:conditions}, respectively.

\subsection{Sequence to decouple all motional errors}


\begin{figure}
    \centering
    \includegraphics[width=3.35in]{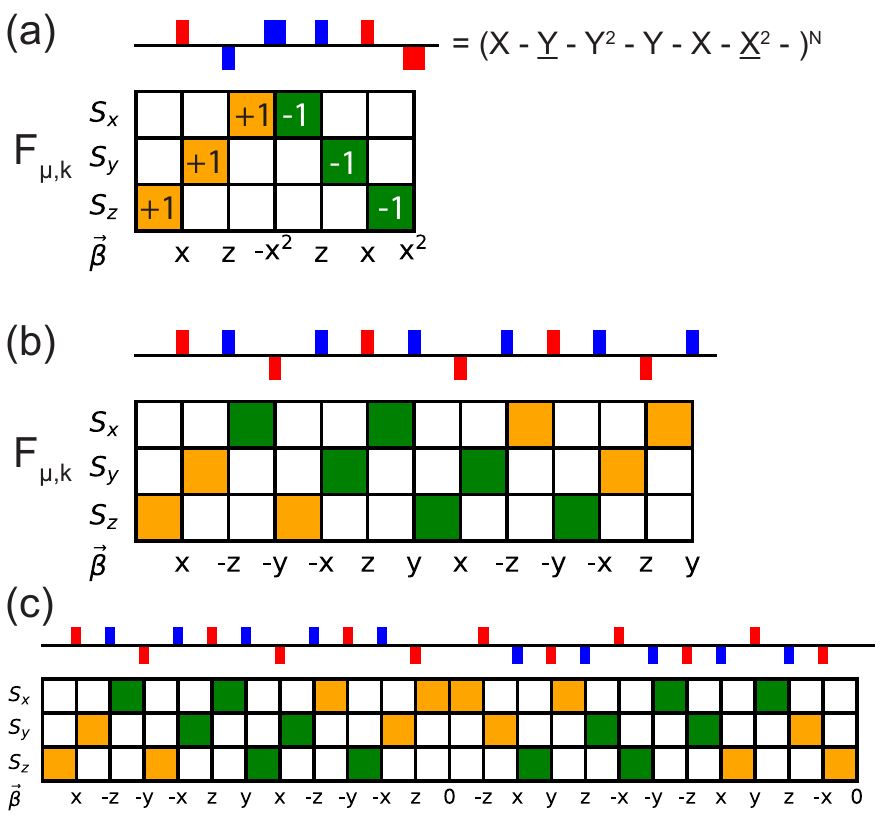}
    \caption{Pulse sequences. In each panel, the top shows the sequence in the lab frame (following Fig. \ref{fig:int_storage}b,c), while the bottom shows $F_{\mu,k}$ in the graphical notation of Ref. \cite{Choi2020}. The $k^{th}$ column gives the value of $F_\mu$ during the $k^{th}$ toggling frame, with +1 (-1) entries shown in orange (green). The very bottom gives the toggling frame rotation axis $\vec{\beta}_k$ connecting the frames.
    (a) Sequence 1.
    (b) Sequence 2. Importantly, the toggling frame Hamiltonian is perfectly periodic: $F_{\mu,k}$ has period $T$ for all $\mu$, $|F_{\mu,k}|$ has period $T/4$ for all $\mu$, and $\vec{\beta{k}}$ has period $T/2$. (c) Sequence 3, which consists of sequence 2 appended with its inverse. Each half maintains the periodicity of sequence 2.
    \label{fig:sequences_SI}}
\end{figure}

In Table \ref{tab:conditions}, we summarize the conditions for the DD sequence to decouple from the effect of atomic motion on various terms. We also include certain conditions from Ref. \cite{Choi2020} that must be satisfied to decouple from (static) disorder and interactions and rotation angle error. The motion-dependent conditions are appended as conditions 4-8.

In Fig. \ref{fig:sequences_SI}, we present diagrams representing $F_{\mu,k}$ and $\vec{\beta}_{k}$ for sequences 1-3 discussed in the main text. The graphical representation follows Ref. \cite{Choi2020}, where sequence 1 is also presented. Ref. \cite{Choi2020} also presents a sequence, ``sequence G", which is conjectured to be the minimum-length sequence that satisfies conditions 1-3 (plus an additional condition relating to errors from finite pulse duration, which we do not re-derive here) with only $\pi/2$ pulses. Our sequence 2 is slightly modified from ``sequence G" by permuting several of the pulses, to retain these characteristics while also making it periodic in all components of $F_{\mu,k}$, $|F_{\mu,k}|$ and $\vec{\beta}_{k}$, which allows the additional conditions to be satisfied as well.

Let us examine sequence 2 in more detail. $F_{\mu,k}$ has one period per sequence repetition for each $\mu$, such that condition 4 can be satisfied if there are an even number of motional periods over the sequence, $\bar{\omega} T = 2 \pi n$ with $n = 2$, 4 or 6 (since there are 12 pulses in the sequence, $n>6$ is equivalent to $12-n$). Similarly, condition 5 can be satisfied if there are an even number of periods of $2\bar{\omega}$ over the sequence, which is the case for any integer $n$. $|F_{\mu,k}|$ has four periods over the sequence, so condition 6 is satisfied for any number of motional periods other than $n=4$. Lastly, $\beta_{\mu,k}$ has two periods over the sequence for each $\mu$. Condition 7 is satisfied for $n=1$,3,4 or 5, and condition 8 is satisfied for $n=2,4,6$.

{\renewcommand{\arraystretch}{2}
\begin{table*}
    \begin{tabular}{|c|l|l|}
        \hline
         No. & Condition & Description \\
         \hline
         1 & $\sum_k F_{\mu,k} = 0$ & Decouples static $S_z^i$ terms \cite{Choi2020} \\
         2 & $\sum_k |F_{\mu,k}| = N/3$ & Decouples dipolar interactions if $J_z=-2J$ \cite{Choi2020}\\
         3 & $\sum_k \beta_{\mu,k} = 0$ & Decouples static rotation angle errors \cite{Choi2020}\\
         4 & $\sum_k e^{-i k \bar{\omega} \tau} F_{\mu,k} = 0$ or $\bar{\omega}\tau = 2 \pi n$ & Decouples spin-motion term $S_z^i x_i(t)$ (\emph{i.e.}, $\Delta$).\\
         5 & $\sum_k e^{-2 i k \bar{\omega} \tau} F_{\mu,k} = 0$ or $\bar{\omega}\tau = \pi n$ & Decouples spin-motion term $S_z^i x_i^2(t)$ (\emph{i.e.}, $\eta'$). \\
         6 & $\sum_k e^{-i k \bar{\omega} \tau} |F_{\mu,k}| = 0$ or $\bar{\omega}\tau = 2 \pi n$ & Decouples spin-motion terms $S_\mu^i S_\mu^j x_i(t)$ (\emph{i.e.}, $J$, $J_z$). \\
         7 & $\sum_k e^{-i k \bar{\omega} \tau} \beta_{\mu,k} = 0$ & Decouples rotation angle errors proportional to $x_i(t)$. \\
         8 & $\sum_k e^{-2 i k \bar{\omega} \tau} \beta_{\mu,k} = 0$ & Decouples rotation angle errors proportional to $x_i^2(t)$. \\
         \hline
    \end{tabular}
    \caption{Conditions for decoupling from static Hamiltonian terms (1,2,3) and spin-motion coupling (4-8). Each condition must be satisfied for all $\mu \in \{x,y,z\}$. \label{tab:conditions}}
\end{table*}
}

Taken together, there is not a choice of $n$ that satisfies all conditions over one repetition of the sequence. However, if we neglect the variation in the rotation angle (assuming microwaves are used to drive the DD sequence), then conditions 4-6 can be met at $n=2$ (this also satisfies condition 8, and would suppress the variation in $\Omega$ if there is no misalignment of the addressing beams to the trap center). The performance of this sequence is shown in Fig. \ref{fig:errors} for various temperatures. Note that this simulation treats the atomic motion as a classical, periodic variation in the Hamiltonian parameters.

If it is necessary to satisfy conditions 7 and 8, the simulation results in Fig. \ref{fig:errors}d suggest that condition 6 could be dropped since the magnitude of the error resulting from variations in $J,J_z$ is considerably smaller than the others. In this case, $n=4$ satisfies conditions 4,5,7 and 8. Alternatively, choosing $n=12$ (one pulse per period) will satisfy 4-6, and also 7 and 8 if condition 3 is also met.

\subsection{Higher order AHT terms}
The framework above can be extended to compute higher-order terms in the Magnus expansion. While the numerical calculations presented in Fig. \ref{fig:errors} capture the influence of terms beyond $\bar{H}^{(0)}$ for the spin hamiltonian, they do not incorporate higher-order terms in the spin-motion coupling, because the motion is treated classically. In fact, conditions 4-8 in Table \ref{tab:conditions} can also be derived from the perspective of constructing a filter function \cite{biercukDynamicalDecouplingSequence2011,degenQuantumSensing2017} to decouple from classical noise. There are no higher-order terms from the motion by itself: the lowest order average Hamiltonian for the harmonic oscillator is exact in the interaction picture. Therefore, the unexplored terms are those arising from spin-motion coupling at higher orders. While these terms can sometimes be important, for example giving rise to geometric phases in trapped ion gates \cite{leibfried2003}, we believe that they are not significant compared to the higher order terms from the pure spin Hamiltonian that dominate the errors in Fig. \ref{fig:errors}, because the magnitude of the spin-motion coupling is small. However, a detailed exploration is left to future work.

\subsection{Active state gates}

Lastly, we consider the active state gates, which includes contributions from $H_{sa}$ and $H_{aa}$. Since the DD sequence is only applied to the storage states, the $\bar{S}$ operators describing the active states do not change with $k$. The toggling frame Hamiltonian for a pair of atoms in the active states surrounded by an array of atoms in the storage states is:

\begin{equation}
    \begin{split}
    \hat{H}_{k,act}(t) &= \sum_{i \in \{1,2\}} \frac{1}{2}\eta_a' m \overline{\omega^2} \hat{x}_i^2(t) \bar{S}_z^i + J_z^{aa}(1-\alpha_z^{aa}\hat{x}_{12}(t)) \bar{S}_z^1 \bar{S}_z^2 + \Delta^{aa}(1-\alpha_\Delta^{aa} \hat{x}_{12}(t)) \\
    &+ \sum_{i \in \{1,2\}} \sum_{j} \left[ \Delta^{sa}(1-\alpha_\Delta^{sa} \hat{x}_{ij}(t))  + J_z^{sa}(1-\alpha_z^{sa} \hat{x}_{ij}(t)) \sum_\mu F_{\mu,k} S_\mu^j \right] \bar{S}_z^i n_j 
    \end{split}
\end{equation}

Here, the index $i$ runs over the pair of sites in the active states, numbered 1 and 2, and $j$ sums over the surrounding sites in the storage states.

Following the discussion above, it is clear that if the period $T$ of the DD sequence is a multiple of $2\pi/\bar{\omega}$, the time-dependent terms with coefficients $\eta_a'$, $\alpha_z^{aa}$, $\alpha_{\Delta}^{aa}$ and $\alpha_\Delta^{sa}$ will vanish. Furthermore, the final $J_z^{sa}$ term will vanish entirely if the conditions $\sum_k F_{\mu,k} = 0$ and $\sum_k e^{i k \bar{\omega} \tau} F_{\mu,k} = 0$ are met for all $\mu$ (conditions 1 and 4 in Table \ref{tab:conditions}). In that case, the remaining terms are:

\begin{equation}
    \bar{H}^{(0)}_{act} = \sum_i \frac{1}{2} \eta_a' m \overline{\omega^2} (2 a_i^\dag a_i + 1)\bar{S}_z^i +J_z^{aa} \bar{S}_z^1 \bar{S}_z^2 + \sum_{ij} (\Delta^{aa} \bar{n}_j + \Delta^{sa} n_j) \bar{S}_z^i
\end{equation}

If we additionally choose $J_z^{aa}$ such that a nonlinear phase of $\pi/2$ is accumulated during $T$, we can return the atoms to the storage states, apply a $\pi$ pulse swapping $\ket{0_s}$ and $\ket{1_s}$, and return to the active states again to realize the same evolution but with $\bar{S}_z^i \rightarrow -\bar{S}_z^i$. The first and third terms will cancel between the two evolution times, but not the second, which leaves the desired interaction. We note that it is not strictly necessary to cancel the third term: its value is known so it could also be incorporated as a single-qubit phase.

\twocolumngrid

\bibliography{test}

\end{document}